# Mitigating Barriers to Public Social Interaction with Meronymous Communication

Nouran Soliman
CSAIL, Massachusetts Institute of Technology
Cambridge, Massachusetts, USA
nouran@mit.edu

Hyeonsu B Kang
Human-Computer Interaction Institute, Carnegie Mellon University
Pittsburgh, Pennsylvania, USA
hyeonsuk@cs.cmu.edu

Matthew Latzke
Allen Institute for AI
Seattle, Washington, USA
mattl@allenai.org

Jonathan Bragg
Allen Institute for AI
Seattle, Washington, USA
jbragg@allenai.org

Joseph Chee Chang
Allen Institute for AI
Seattle, Washington, USA
josephc@allenai.org

Amy X. Zhang
CSE, University of Washington
Seattle, Washington, USA
axz@cs.uw.edu

David R Karger
CSAIL, Massachusetts Institute of Technology
Cambridge, Massachusetts, USA
karger@mit.edu

## ABSTRACT

In communities with social hierarchies, fear of judgment can discourage communication. While anonymity may alleviate some social pressure, fully anonymous spaces enable toxic behavior and hide the social context that motivates people to participate and helps them tailor their communication. We explore a design space of *meronymous communication*, where people can reveal carefully chosen *aspects* of their identity and also leverage trusted *endorsers* to gain credibility. We implemented these ideas in a system for scholars to meronymously seek and receive paper recommendations on Twitter and Mastodon. A formative study with 20 scholars confirmed that scholars see benefits to participating but are deterred due to social anxiety. From a month-long public deployment, we found that with meronymity, junior scholars could comfortably ask "newbie" questions and get responses from senior scholars who they normally found intimidating. Responses were also tailored to the aspects about themselves that junior scholars chose to reveal.

## CCS CONCEPTS

• **Human-centered computing** → **Social networking sites**.

## KEYWORDS

Q&A; Social Recommendation; Online Safety; Online Communities; Social Media; Partial Anonymity; Self-Presentation; Identity; Self-Disclosure

## 1 INTRODUCTION

Speaking up can be frightening. When people communicate with others, they may worry (justifiably or not) about how their words will be received and how the speaker will be judged. Beyond causing stress, these concerns can deter people from speaking up at all, preventing valuable sharing of information. Speaking up *in public* can amplify these concerns as one faces the judgment of many individuals. And while anyone may fear to speak, in communities with strong social hierarchies, lower status individuals may particularly fear the judgment of higher status ones and be particularly reluctant to speak. As a result, public spaces can become skewed towards the voices of more senior or high status members of a community [34] or those with other privileges that shield them from social ramifications, in some cases stifling the growth of communities [33]. Those who can speak up more can also sometimes reap personal benefits, producing a rich-get-richer effect.

This phenomenon is easy to recognize in the academic community, a setting widely known to be hierarchical [11], with many signals for conveying status and prestige [9]. In addition, the importance of one's social standing and social network [53, 54] for promotion and the overall competitive environment makes almost every public social interaction potentially high stakes. As a result, online public social spaces where academics converse can be dominated by the voices of senior scholars [76], leaving junior scholars to turn to smaller and more private spaces for social exchange. Past research indicates that compared to professors, students are less inclined to use Twitter for conference participation and learning activities, but more for non-academic reasons [42, 44, 52, 92]. As a concrete example, we confirmed via a formative study with 20 scholars that they, particularly those early in their careers, perceive interactions with senior scholars as high-stakes. As a result they refrain from speaking up in public settings such as Twitter (now called X) due to a fear of being perceived as uninformed. This has serious negative ramifications for these junior scholars; research has shown that engaging in public spaces like Twitter can boost one's visibility and even garner more citations for one's work [43].

One solution used in online settings to help people feel safer when posting is anonymity. Hiding one's identity can indeed alleviate social pressure but full anonymity obscures important social context that helps responders tailor their responses. In addition, people tend to engage less with anonymous posts [56], as they are more motivated to respond to people when they know certain aspects about them, for instance, if they share some background in common or they are part of the same community. Anonymity can also be *too* successful in alleviating social pressure, enabling people to troll and harass others without consequences or engage in hateful speech, eventually making some fully anonymous social spaces highly toxic (*e.g.,* the Economics Job Market Rumors forum [21, 98]).

In this work, we consider ways to obscure people's *exact* identity to alleviate social pressure in public discourse, while still *verifiably* disclosing relevant *aspects* of their identities and using *endorsement* to preserve *accountability* to foster an active, high quality, and positive public space. We present **meronymity** (from the Greek "mero-", partial, and "nym", name), a design paradigm where people can make nuanced decisions around what meronym to reveal about themselves and to whom for specific threads of discussion. A **meronym** is a trustworthy description of an individual, composed of a number of aspects of their identity (identity signals), that provides pertinent information about them without revealing their identity.

For example, someone might wish to post as *"a PhD student in an HCI program at a research institution and who has 1–5 publications at CSCW and CHI."* This meronym signals to potential responders that the post is from someone who is already embedded in the HCI community and who will understand the jargon of the field in responses. The revealed aspects could even be personalized to specific receivers to highlight, for instance, commonalities between the two people to encourage a better response. A person might also be willing to share their full identity privately with a few trusted individuals; those individuals could then serve as public *endorsers* of the poster or help to boost the poster's public message.

We implemented these design ideas into a system called LiTweeture that is geared towards Q&A for academic literature recommendations on Twitter and Mastodon. We chose this topic as our formative study found that junior scholars find social literature recommendations useful when they encounter them publicly, yet are typically reluctant to participate. We chose to build on top of existing social media platforms instead of creating a new social space dedicated to academic discourse, as our formative study found that these platforms are the primary places where academics converse publicly online, and building on them enables us to observe more realistic usage [30].

We conducted a month-long public field study. To catalyze participation, we recruited 13 junior scholars to regularly use the system and post questions and answers. Participants expressed that they were more comfortable asking *"newbie"* and other *"embarrassing"* questions on LiTweeture that otherwise would have been left unasked, and were also comfortable posing them to people they didn't personally know, including more senior scholars for whom they normally would be too intimidated to contact. Participants also enjoyed having the flexibility to choose how much to reveal about themselves, and these signals helped give responders context on appropriate answers. By leveraging their existing relationships to get endorsements and re-posts, participants were able to get their questions in front of more social media users and were encouraged by the responses they received.

We conclude by discussing the broader applicability of meronymity as a facilitative feature in various online interactions, extending beyond the academic sphere and literature recommendation into public discourse at large.

In summary, the main contributions of our work are:

- A deeper understanding of motivations of and social barriers for scholars in online public engagement with seniors, impressions of scholars on partial anonymity, and how the design of meronymous interactions could be informed. We achieved this by holding a formative study with 20 scholars.
- The design of a generic meronymity framework with affordances that enable users to harness some of the benefits of public interaction while minimizing the pitfalls of anonymity. This step was informed by previous literature and our formative study results.
- A system, LiTweeture, that incorporates our proposed meronymity framework in the context of public research paper recommendations within academic communities on Twitter (now called X) and Mastodon. During this step, design iterations and pilot studies were held to refine the system, its affordances, and identity descriptions.
- Empirical understanding of how scholars would use and perceive our proposed affordances. We achieved this step by deploying LiTweeture in a one-month field study with 13 scholars.
- A discussion of the broader applicability of meronymity for facilitating discourse beyond the academic sphere.

## 2 RELATED WORK

### 2.1 Barriers to Public Engagement in Online Communities and Approaches for Mitigation

Online communities such as Reddit[1], Quora[2], and Twitter[3] have emerged as highly popular platforms for public engagement. These communities often cater to information needs that traditional search environments may find challenging to fulfill, such as personal and health-related questions. It is not just the ability to draw on the lived experiences of others in response to unique questions that makes these platforms appealing, but their trustworthiness and social engagement factors that set them apart from search engines [66, 70]. Notably, collective contributions from diverse users can result in comprehensive, reusable answers [32, 55].

---
[1] https://www.reddit.com/
[2] https://www.quora.com/
[3] https://twitter.com/home

However public engagement in online communities faces several challenges. One such issue is the imbalanced attention that questions receive, with popular users often netting more responses, showcasing a "rich-get-richer" effect [73]. Alongside such operational challenges are social issues tied to seeking recommendations from online communities. Users may hesitate to ask for help due to reasons such as fear of reciprocity obligations [90], need for personal information disclosure [8, 95], reputational concerns [14, 19, 26, 50, 51, 96], or simply the unease of bothering others [8, 15].

The situation is further complicated by the existence of marginalization within these online communities, which acts as a significant deterrent for user engagement. Marginalized users are less likely to identify with the community or feel compelled to contribute [39], both factors that increase users' commitment to participation [71, 79]. For example, prior research shows extensive gender-based marginalization on Wikipedia [64], Github [89], StackOverflow [27], and in open source software (OSS) participation in general [25, 63], with additional research identifying its sources as stereotyping or unwelcoming language [71, 91] and biases in the perception of women's technical competencies [86]. The research extends to existing community hierarchies (*e.g.,* newcomers vs. senior members in a community on Wikipedia or in OSS [80, 102]) and power dynamics that discourage users on the lower rungs of perceived hierarchies from participating in discussions, or prevent them altogether from raising challenging topics for discussion due to fear of hurting relationships [5].

A study of online community tools and platforms highlights their important role in perpetuating or even amplifying marginalization [63]. To this end, efforts to lower barriers have led to the proposal of improved tools, such as the development of a portal for streamlining on-boarding on Wikipedia [81], recommender systems that help find mentors [10] or code reviewers [100], and ephemeral platforms [56]. However, these tools showed limited success, increasing newcomers' participation in the short-term [23], but struggling with sustaining long-term engagement [46] or addressing the challenges that senior community members face [3]. And, despite these efforts, the burden of mitigating barriers to participation often falls upon the marginalized group [87], highlighting areas for potential improvement in platform design and tools.

## 2.2 Motivation and Challenges in Public Scholarly Interactions

The challenges arising from social considerations are especially pronounced in public scholarly interactions where potential professional risks stemming from rigid hierarchies and competition loom large [1, 29]. However, a growing number of academics are drawn to social networking platforms like Twitter, Academia.edu[4], and ResearchGate[5] for the accessibility and visibility they offer to their own and new research [29, 72, 83, 85, 97]. In fact, such platforms can provide significant benefits [59] to those who participate, such as enhanced scholarly impact exemplified by increased citations (*e.g.,* having any tweets of a publication leads to four more citations [43]) and expanded visibility of their work (*e.g.,* links to publications garner significant retweets and likes [24]). Another study has shown that online social networks in higher education environments could affect the construction of students' professional identity through online social capital as a mediator [38, 59]. While junior scholars acknowledge these benefits [83, 103], their potential vulnerability to social ramifications that may affect career advancements can deter their active participation [29, 36]. As a result, participation may be dominated by faculty, who produce significantly larger proportions of scholarly tweets compared to non-faculty [76]. Furthermore, deeper issues of disparities, such as gender inequality, also exist within academic-public social interaction. Despite women's high engagement with social media tools, their participation in scholarly communication on social media is surprisingly low [45, 77, 78].

## 2.3 Self Disclosure and Audience Management

Self-disclosure in social media, a nuanced and multi-faceted phenomenon, has been the subject of extensive research due to its significant psychological and social implications. The motivation for self-disclosure often stems from the desire for social support, connectedness, and relationship development. Studies [57, 94] have underscored this aspect, emphasizing how sharing personal information online can foster a sense of community and emotional closeness. Yet, this practice is not without its pitfalls. The risks involved, such as loss of privacy, potential stigma, and exposure to harassment, necessitate a careful balancing act [20, 57]. The decision to disclose is influenced by various factors, including the level of anonymity afforded by the platform, the nature of the audience (whether friends, family, or strangers), the intimacy of the content being shared, the prevailing norms of the platform, and the positivity or negativity of the message [6, 57]. For instance, users are often more willing to share sensitive information under the cloak of anonymity or with weaker social ties, as these conditions reduce perceived risks.

The audience plays a critical role in shaping online self-disclosure. People tend to be more reserved in their sharing when faced with a "collapsed" audience; a mix of friends, family, colleagues, and acquaintances all in the same virtual space [7, 13, 61, 65, 93]. This phenomenon has led users to adopt various audience management strategies to navigate these complex social webs. These strategies include creating separate social media accounts for different aspects of one's identity, using specific platforms known for catering to particular audience types, and curating content to suit each audience group [17, 82, 99].

However, the implementation of these strategies comes with its own set of challenges. Providing users with flexible access controls that are not overly burdensome, while at the same time safeguarding the platform from toxic behaviors that anonymity might enable, is a significant hurdle [49, 88, 101].

In this context, the concept of "meronymity," as introduced in this project, offers a promising avenue for exploration. It seeks to provide a middle ground where users can disclose aspects of their identity selectively, ensuring credibility while maintaining a degree of anonymity. This approach could effectively address the challenges of audience management and self-disclosure by enabling users, especially in hierarchical environments like academia, to communicate in a way that is both safe and authentic. This could fill a significant gap in current research, which has not fully explored the balance between credibility and privacy in the context of partial anonymity.

---

[4]https://www.academia.edu/
[5]https://www.researchgate.net/

## 2.4 Identity Presentation in Online Communities: An Overview

In the realm of online communities, how individuals choose to present their identity is a fundamental aspect that shapes interactions and communication. The digital world offers a spectrum of identity options, from complete anonymity to fully identifiable profiles, each with its own set of implications for user behavior and community dynamics. Here we focus exploration on three primary modes of identity presentation: anonymity, pseudonymity, and partial anonymity. Each mode offers unique benefits and challenges, influencing the nature of online discourse and social engagement. A deeper understanding of these identity mechanisms is essential to grasp the complexities of digital interactions and community building, and to inspire new models of identity representation that could create safer and protected spaces.

*2.4.1 Anonymity in Online Communities.* The challenges of public social interaction have led researchers to consider the potential benefits of anonymity in fostering online help-seeking behavior. The concept of anonymity in online platforms has been a subject of extensive research, with studies revealing a complex array of impacts on user behavior. Anonymity, characterized by the absence of identifiable information, allows users to engage in discussions without the fear of personal judgment or social repercussions. Studies indicate that the ability to pose questions anonymously significantly mitigates the sense of accumulating social debt without compromising the quality and quantity of responses [31, 56]. This freedom can lead to increased self-disclosure, especially in sensitive contexts such as online health communities [40]. For instance, individuals dealing with stigmatized illnesses are more likely to seek and offer support in an anonymous setting, a phenomenon further explored in previous research [2, 68].

However, the shield of anonymity can also embolden users to engage in negative behaviors. The concept of the online disinhibition effect [84] suggests that anonymity can lead to a reduction in social inhibitions, paving the way for toxic behaviors like trolling [47]. Research suggests it can diminish users' sense of accountability and promote deviant behaviors within online communities [16, 84]. Other studies [35, 48] further explore the darker side of anonymity, demonstrating how it can foster cyberbullying and deceit.

In response, an emerging area of interest extends beyond the binary decision of anonymous or public visibility, questioning the degree to which anonymity protection is necessary. Comparative analysis of content on anonymous (*e.g.,* Whisper[6]) versus non-anonymous (*e.g.,* Twitter) platforms indicates varying levels of anonymity sensitivity across different content categories, ranging from low (*e.g.,* Humor) to high (*e.g.,* Meet-ups) [12]. Another line of research has also delved into more nuanced anonymity models, such as tie-based anonymity. For example, a study [57] describes how platforms like Secret and Mimi offer a form of anonymity that is contextualized within one's social network, providing a blend of anonymous expression and social context. Despite this innovation, challenges persist, including the risk of bullying and negative behavior within these semi-anonymous networks [41]. This suggests that not all content requires equivalent anonymity protection and that platforms could potentially better support different genres of public discourse by supporting different modes of identity presentations.

*2.4.2 Pseudonymity and Its Dynamics.* As an alternative mode to full anonymity and full identity disclosure, studies on pseudonymity also contribute to this discourse. Pseudonymity, as distinguished from full anonymity, involves the use of consistent, yet non-identifying online personas. This form of identity management allows individuals to build and maintain an online reputation and relationships [18]. It offers a degree of privacy while fostering a sense of community and accountability not typically found in fully anonymous environments.

According to research, pseudonymous and anonymous users engage in Q&A interactions of equal quality and quantity [31]; however, it was also noted that anonymous questions are more prone to trolling which can degrade answer quality and have a negative impact on archival value. The benefits of pseudonymity over anonymity are significant. While it maintains a level of privacy, it also discourages trolling and encourages the accrual of social capital in online communities. Users can engage in more prolonged and meaningful interactions, building trust and credibility over time [49]. However, the persistent nature of pseudonyms also means that negative reputations or histories can follow a user, making it challenging to escape past interactions or disclose only relevant context-specific identity aspects in a given discourse [49].

To navigate the challenges of pseudonymity, users often employ strategies such as creating throwaway accounts [49], especially on platforms like Reddit. These temporary pseudonyms allow users to engage in discussions or disclose sensitive information without the burden of their established online persona. This practice, while addressing some of the limitations of pseudonymity, introduces complexities in managing multiple identities and the loss of accumulated reputation associated with a primary account. These insights indicate the importance of context-specific anonymity and identity presentation in online social interactions, and point to a gap in literature on partially anonymous interactions beyond alternative identity presentation designs such as pseudonymous interaction.

*2.4.3 Meronymity: A novel approach balancing Credibility and Privacy.* In this work, we formalize and present **meronymity**, a design paradigm that aims to address the challenges and limitations inherent in anonymity and pseudonymity. Meronymity proposes a model where users can selectively reveal different facets and social cues of their identity, allowing for a more dynamic and flexible presentation of self in online communities. This approach could offer a solution to the trade-offs between the risks of full anonymity, such as lack of accountability and negativity, and the constraints of persistent identities in pseudonymity. This approach can also enhance the quality of interactions by providing sufficient context to participants while preserving a degree of privacy.

---
[6]https://whisper.sh/

Meronymity could enable users to navigate different online contexts more fluidly, revealing aspects of their identity that are relevant and beneficial to the interaction while retaining the ability to protect their overall privacy. This approach holds the potential to enrich online discourse by allowing users to bring in relevant aspects of their identity, expertise, or experience when needed, without the burden of a singular, unchanging online persona. This facilitates avoiding issues like context collapse and juggling multiple throwaway accounts for privacy. Meronymity could also enable platform understanding of user activity patterns to better support people and content moderation while protecting anonymity. We incorporate our proposed meronymity model to LiTweeture, a system that supports users in making nuanced decisions around what aspects to reveal about themselves and to whom for a specific public help-seeking context, and indirect third-party endorsement.

Early explorations around the idea of sharing social cues about one's self online were conducted by prior research. Research drawing on the Social Identity Model of Deindividuation Effects (SIDE) [75] has shown that revealing specific identity cues, such as political affiliation or academic credentials, can positively impact online discussions and credibility. For example, previous studies [28, 69] found that disclosing certain credentials or affiliations can increase influence and credibility in online interactions, suggesting that partial anonymity can enhance the quality of discourse without compromising individual privacy. The concept of group anonymity, as explored in platforms like SecurePost [67], offers an intriguing application of partial anonymity. In such systems, group affiliations are revealed, but individual identities remain concealed. This approach allows for the preservation of individual privacy while providing a contextual background that can foster trust and credibility within the community.

## 3 FORMATIVE STUDY

Choosing the academic community as a test-bed for implementing meronymity, we conducted a formative study by interviewing 20 scholars to gain a deeper understanding of the nature of their public academic and help seeking interactions with senior figures, especially on social platforms like Twitter. Our Study received Internal Review Board approval.

### 3.1 Objectives & Research Questions

In this study, we aimed to comprehensively understand the social impediments inherent in what our participants characterized as public *"high-stakes"* engagements between scholars and senior figures. We particularly focused on the motivations of and obstacles faced by scholars in paper recommendation exchanges and discussions in the three following parts of the activity: asking for, sharing, and finding/encountering research paper recommendations on Twitter. Additionally, we explored scholars' feelings and perceptions around creating and engaging with partially anonymous threads publicly for exchanging research papers. Our main objective was to leverage the study results to inform and refine our design ideas and framework of meronymous interactions when publicly seeking assistance or engaging in discourse within the academic community, and more broadly other communities with parallel social dynamics.

Our research questions encompassed:

- How do scholars get access to (encounter) paper recommendations online?
- How do scholars exchange paper recommendations with their community online?
- What are some of the challenges scholars face when exchanging paper recommendations publicly?
- What do scholars feel about engaging with their community using partial anonymity?

### 3.2 Procedure

We organized a 1-hour interview session with scholars, asking them about their experiences and thoughts on exchanging research paper recommendations with peers. We were especially interested in public interactions, so we first asked about their experiences and feelings when encountering, discussing, sharing and asking for research paper recommendations on platforms like Twitter. We aimed to understand what drives or discourages them from such engagements. We then discussed the role of partial anonymity on platforms like Twitter and how it might influence their interactions. The session concluded with a demographics survey.

To recruit participants, we spread the word through personal contacts, academic mailing lists, HCI groups on Meta, and posts on the personal Twitter and Meta accounts of our team and colleagues. We ensured our participants were active researchers and above the age of 18. After completing the interview, participants received a compensation of $20/hour (USD) via PayPal.

For our data analysis, we used an inductive thematic approach. The primary author conducted the analysis, refining the themes in consultation with co-authors to ensure consensus was achieved. We did not calculate inter-rater reliability because the codes were developed as part of a thematic analysis to yield concepts and themes [62]. In the sections that follow, we share insights from our participants, with quotes attributed using the format "P-participant number."

### 3.3 Participants

We interviewed 20 scholars (5 women, 13 men, 1 non-binary, and 1 queer). Their academic experiences varied: 1 undergraduate, 13 graduate students, 1 post-doc, and 5 industry researchers. In terms of seniority, 8 were juniors (undergraduates and early graduate students) while 12 were seniors (senior graduate students or industry researchers and post docs). Their ages ranged from 21 to 54, with 2 between 21-24, 15 between 25-34, 1 between 35-44, and 2 between 45-54. Every participant was from the USA. They specialized in various fields: 11 in Human-Computer Interaction, 4 in Natural Language Processing, 3 in Machine Learning, 1 in Computer Graphics, and 1 in Cryptography.

### 3.4 Results

*3.4.1 Participants regarded Twitter as a key resource for high-quality paper recommendations.* Participants overwhelmingly viewed Twitter as an important platform for procuring high-quality research paper recommendations from the research community. A significant number of participants shared that they habitually turned to Twitter as a primary source for the latest research findings, especially emergent ones. As P-15 succinctly put it, *"I like to look a lot on Twitter...I do go there to get recommendations from people that I follow on Twitter and I curate it over time."* Expanding on this, many have honed their Twitter networks by actively following renowned experts in their domain to ensure they remain updated with the most relevant papers. P-16's statement, *"I try to follow all the people that are relevant to my research on Twitter, and I see a lot of the paper recommendations there"* further reflects this trend.

The immediacy of Twitter was a recurring point of emphasis, where participants noted the spontaneity with which new research is shared, often moments after its release. P-19 captured this sentiment, explaining, *"Nowadays it's mostly from Twitter, so you are inevitably seeing a lot of papers where you're just scrolling down Twitter and then at that time I will see some of the latest work, which I believe others don't see cause you know, the tweeting time, just like one hour ago or something like that."*

The value participants found in Twitter's recommendations was clear. P-17 emphasized the depth of insights they gained from Twitter threads shared by members of their community, stating that *"single paper threads...are really valuable because they're much better than an abstract in getting a feel."* This sentiment was echoed by several participants who expressed that they frequently bookmark these recommendations. P-11 reflected a similar opinion, *"I do wanna save these papers because they are like collective wisdom and they're from people who are actively working in these fields and they are like actual good recommendations."* Moreover, participants also highlighted the serendipity and diversity of papers they discovered on Twitter, with P-3 recounting, *"A little while ago...a professor posted their top 10 papers in cognitive science. And I found that list super useful and really interesting. And it allowed me to dig into some literature that I'd never seen before that was actually really great."* Twitter's role as a pivotal hub for accessing and sharing high-quality research paper recommendations among academic communities was unequivocally evident from the participants' experiences and insights.

*3.4.2 Participants showed a reluctance to ask for, discuss, and share paper recommendations publicly on Twitter despite the perceived benefits.* Participants elucidated their experiences and perspectives on exchanging research paper recommendations, particularly within public platforms such as Twitter. While participants acknowledged the value of engaging within the Twitter community and sharing recommendations, a predominant inclination was to ask for and disseminate papers privately or within confined groups through mediums such as Slack, Twitter DMs, or email, often targeting collaborators, advisors, or close acquaintances. Several participants articulated reservations about engaging publicly due to a variety of concerns. P-15 voiced apprehensions about maintaining a public persona, noting, *"I worry about my image on social media."* Similar sentiments were echoed by others, such as P-11, who expressed, *"I feel like my judgment of the field will be judged by other people."* A pervasive sentiment revolved around fear of public scrutiny and judgment, as highlighted by P-20's statement about reluctance to recommend papers that might *"reveal their lack of knowledge."* This apprehension was further encapsulated by P-15's reluctance to share nascent thoughts: *"I don't wanna be contributing...something that I thought just for two seconds."*

Notably, the stature of senior researchers loomed large for some, leading to a hesitancy in contributing to discourse for fear of appearing uninformed. For instance, one participant was less inclined to contribute *"obvious"* or older recommendations to avoid *"looking ignorant in front of many senior researchers."* Furthermore, the notion of "context collapse" [7, 13, 61, 65] emerged as a constraint, where individuals must navigate and integrate various social roles and different audiences in a single online space leading to a blending or "collapse" of multiple social contexts into one. One participant voiced this concern by stating: *"I definitely don't really engage...because of the context collapse problem"* (P-13). Nonetheless, participants identified certain conditions fostering public interactions, predominantly hinged on the strength of social ties. This was stated by multiple participants. For example, P-16 stated, *"If it's like somebody I know...I'm probably more likely to reply,"* whereas another participant emphasized a higher level of closeness to incentive them to engage *"if it's a friend or someone [they] have a close working relationship with"* (P-17).

When it came to seeking recommendations, while participants recognized the potential of Twitter to *"gather the crowdsourced opinion on a set of papers"* (P-19), there existed a pronounced hesitancy. P-15 highlighted the need to balance requests, expressing, *"I don't want just all the time to ask for help."* Other participants felt a *"reluctance to publicly ask fundamental questions"* to evade potential embarrassment, and one participant perceived Twitter inquiries as less efficient, akin to *"gambling"* without certainty of a constructive response. These findings underscore the complex interplay of public engagement, perceived image, social challenges, and the value of recommendations in digital academic spaces.

*3.4.3 Participants indicated a preference for engaging with partially anonymous posts over fully anonymous ones on Twitter.* Scholars were consulted regarding their experiences and perspectives on exchanging research paper recommendations on platforms such as Twitter, especially focusing on public interactions. An intriguing insight emerged from the discussions: participants revealed a distinct preference for engaging with posts that offer partial anonymity over those that are fully anonymous.

While participants acknowledged the comfort that complete anonymity brings in reducing social anxiety around public posting, they expressed reluctance to engage with wholly anonymous posts due to mistrust issues, which corroborates previous work [56]. For instance, P-15 mentioned their own preference to anonymously contribute publicly in order to *"avoid appearing superficial or trivial."* But they then expressed reluctance to respond to other people's anonymous requests, questioning the legitimacy and intent behind such posts: *"I wouldn't be interested in responding to an anonymous thread... Is this a quiz? A test? I don't want to leave a misguided recommendation."* This sentiment was echoed by others, with P-17 emphasizing a willingness to interact only if these posts are associated with trusted entities, suggesting the importance of established credibility in the digital space.

The notion of partial anonymity, however, where certain "identity signals" are disclosed, seemed to find favor among participants.

These signals, participants argued, not only achieved the same impact of full anonymity by reducing the social anxiety associated with public discourse but also lent a semblance of genuineness to the interactions. For instance, one participant felt that these identity markers would aid in tailoring recommendations and fostering a sense of connection instead of feeling that they were encountering *"a random request"*, thereby giving the interaction more weight and purpose.

### 3.5 Summary

Our study finds Twitter vital for scholars to discover quality research and engage with the scientific community. Academics use it for latest findings, yet many are cautious about public engagement due to image concerns and fear of scrutiny. While anonymity eases social anxiety, there's a preference for partial anonymity in posts, balancing privacy with credibility.

Affirming the social challenges encountered by scholars in academic public interactions with seniors, our study insights further motivated the design of a partially anonymous academic platform where scholars would be enabled to selectively reveal aspects of their identity when engaging publicly with their community. Exploring what kinds of identity descriptors would be motivating and credible for scholars in this setting was another question to focus on based on the feedback collected in this study. Additionally, one of the main design choices informed by this study was to build our system on top of Twitter instead of creating a new social space dedicated to academic discourse [30] as Twitter was the primary place where scholars followed paper recommendations, announcements, and discussions online. This choice would enable us to observe more realistic usage, and enrich our evaluation experiment design due to leveraging the existing communities and social connections.

## 4 MERONYMOUS COMMUNICATION

### 4.1 Design Goals

Informed by established literature and preliminary outcomes discussed in our formative study, our research aims to delineate design objectives pivotal to facilitating a pro-social online discussion environment that mitigates barriers to public communication. We begin by outlining some high level concepts and design goals and the relationships between them that inform the design of our system. Here we italicize the concepts that we will address throughout the paper.

We have already introduced the concept of *meronymity*. A *meronymous* post is associated with a *meronym*, a chosen partial description composed of a number of aspects of the poster's identity (identity signals). Several of our design goals center on meronymity. First, of course, is to permit people to make and receive meronymous posts (denoted **DG1**). Recall, however, that the specific hope for meronymity is to overcome two drawbacks of anonymity: that people are not motivated to respond to anonymous posts, and that if they do respond, they don't know how to tailor their response to the poster. Therefore, we aim for our meronymity design to include those descriptors that are necessary to encourage *engagement* with posts (**DG2**) and to tailor their response to the poster (**DG3**).

The meronymity affordance will only be effective if people can trust them; therefore we incorporate the goal of *verifiability* (**DG4**).

Finally, since meronymity creates similar opportunities as anonymity of harassment or toxicity without consequences, we add a design goal of *accountability* (**DG5**)—ensuring that someone who misbehaves can be called to account.

Individuals with few connections to the community or low social capital may have trouble getting their questions noticed by people who can answer them. This requires us to consider mechanisms for *promotion* (**DG6**) of content to a wider audience.

### 4.2 Meronymity Model and Design

In our proposed meronymity framework, we delineate five primary stakeholders: the Poster, Potential Responders, Desired Responders, Endorser, and Helpers (Fig.1).

- Poster: This entity initiates the communication, typically driven by a need for assistance or information.
- Potential Responders: This broad category embodies the entirety of the community, encompassing both those who may and may not possess the requisite knowledge or experience to address the Poster's query.
- Desired Responders: A specialized subset of the Potential Responders, these individuals are uniquely equipped with the pertinent knowledge or insights to aptly address the Poster's inquiry. It is noteworthy that the relational dynamics between the Desired Responders and the Poster can span from personal acquaintances to complete strangers.
- Endorser: This is an individual who is called upon due to their high level of credibility, reputation, or network, who possesses a personal rapport with the Poster. Their role is to lend credibility to the Poster by vouching for their authenticity.
- Helpers: These are individuals with a direct personal link to the Poster and can play a role in amplifying the reach of the Poster's inquiry, helping it garner attention within the community.

The following subsections present additional design details regarding identity information sharing and interactions amongst these stakeholders.

*4.2.1 Public Sharing of Verified Identity Signals.* In advancing our design goals, we propose equipping Posters with tools to selectively present facets of their identity in public communications (Fig. 1, A). This strategic obfuscation and precise control over identity disclosure enables meronymous posting (DG1). Identity indicators that could enhance credibility, add context, and foster meaningful engagement include: their association and rank within a defined community (*e.g.,* a junior researcher in HCI), their expertise relative to the subject in question (*e.g.,* a seasoned researcher with numerous publications at CHI), institutional affiliations (*e.g.,* a doctoral candidate affiliated with [specific institution or corporation]), and relational ties to Desired Responders (*e.g.,* direct or secondary connections they have in common). Conveying such indicators not only augments the perceived trustworthiness and authenticity of the Poster (DG2) but also equips Responders with an enriched relational and contextual perspective. This helps them understand the Poster's expectations, thereby incentivizing tailored, meaningful interactions while keeping the Poster's identity anonymous (DG3).

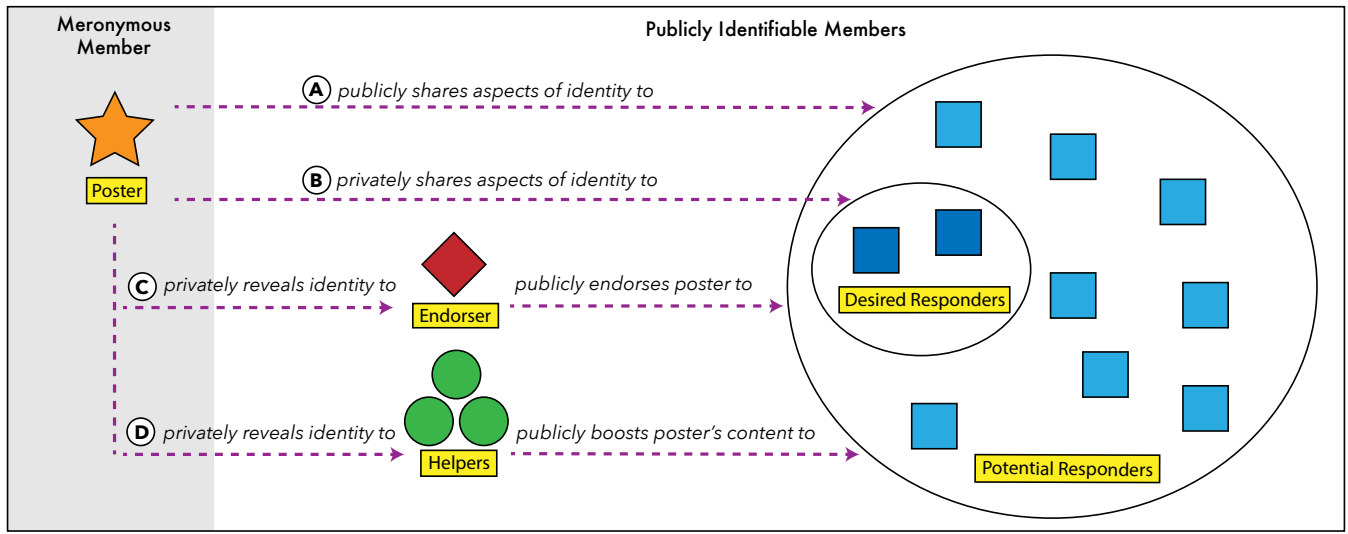

**Figure 1: Identity/Content Sharing in our proposed Meronymity Model.** This diagram delineates the roles and interactions among five key stakeholders. The 'Poster' initiates communication for assistance, choosing varying anonymity levels when sharing identity: (A) In full or partial using aspects of identity (meronym) publicly to Potential Responders, (B) In full or partial using personalized aspects of identity privately to Desired Responders for targeted communication, (C) In full privately to an Endorser for public endorsement, leveraging the Endorser's identity to enhance credibility, (D) In full privately to Helpers who amplify the content to Potential Responders.

Nonetheless, an inherent risk persists: without verification, the anonymity could be exploited, leading to potential misuse, harassment, or spamming. It is imperative, therefore, to embed robust identity verification mechanisms. One strategy involves using data about the Poster from trusted entities, known for rigorous and sometimes manual identity vetting processes, akin to Twitter's former identity verification process for getting a blue check. Upon successful verification, an expanded array of identity indicators could be curated, rooted in the Poster's activities within the verifying platform (DG4). Another example of identity verification is leveraging affiliations of the Poster to some entities that could reveal something about them. For example, using work or school ".edu" emails is one way, albeit unreliable, of verifying that the Poster attends an institution.

*4.2.2 Personalization of Identity Signals in Direct Messaging.* Multiple challenges related to posts being overlooked by Desired Responders have informed our second design consideration. Within our preliminary study, participants highlighted the unpredictable nature of platforms like Twitter, wherein posts can transiently appear and subsequently vanish while scrolling, leading to limited engagement if not seen at the right time. A sentiment echoed by several participants was a hesitancy to pose questions on social media due to past experiences of non-responsiveness. A possible explanation for this, as speculated by participants and supported by prior research, may lie in the platform's filtering algorithms [22] which might prevent Desired Responders from viewing the post at an opportune moment. Additionally, research indicates that social media platforms can engender a diffusion of responsibility [4, 60], resulting in potential Responders not responding even when presented with relevant posts.

To counteract this challenge, we suggest the proactive notification of a select group of Desired Responders about the post in question (DG3). However, this does not guarantee engagement. There remains the possibility that Responders, especially those with significant domain expertise, may not feel sufficiently incentivized to engage, given the likely more generic identity cues the Poster shares publicly.

To remedy this, we introduce a provision allowing Posters to selectively disclose more personalized identity markers exclusively to their Desired Responders (DG2), (Fig. 1, B). This may encompass more detailed self-identifiers (*e.g.,* "a PhD student in HCI affiliated with [specific institution], previously collaborating with [specific researcher]"), demonstrable relational ties (*e.g.,* "previously collaborated on a project with you"), or possibly even the Poster's full identity. By privately sharing enhanced identification details with the Responder, we aim to strike a balance: on one hand, enhancing the likelihood of engagement by capitalizing on shared professional or social networks, and on the other, maintaining public anonymity for the Poster.

*4.2.3 Endorsement by High-Status Community Member.* To enrich identity presentation options for the Poster (DG1), bolster their credibility (DG2), incentivize Desired Responders to engage (DG2) with them, and introduce a degree of verification (DG4), accountability (DG5) and promotion (DG6), the role of an Endorser becomes pivotal. We propose a mechanism wherein the Poster can associate themselves with a recognized, reputable senior individual who will publicly endorse the poster. In this setup, the Poster would confidentially disclose their identity to the Endorser. Subsequently, the Endorser acts as an intermediary, strengthening the connection between the Poster and Potential Responders (Fig. 1, C). This allows Posters to accentuate their posts by highlighting their endorsement by a recognized figure, further enhancing the robustness of their identity presentation. Another crucial role that Endorsement serves is ensuring accountability of Posters. By employing this relationship, any inappropriate actions or messages from the Poster can be traced back to their Endorser, establishing a clear line of responsibility. The Endorser, in turn, would bear the implications of having vouched for the individual in question or cooperate to prohibit this kind of behavior (DG5).

Note that unlike boosts in Mastodon or "likes" on Twitter, endorsements are granted to a *poster* rather than a post.

Once a Poster has been endorsed, they can make as many posts as they want under the (single) endorsement by their endorser. We made this choice to keep the burden on endorsers low (since we are essentially asking them for a "donation" of social capital, we don't want to make it hard for them to donate) and to avoid delays that would accumulate waiting for an endorser to endorse a just-authored post.

*4.2.4 Leveraging Trusted Individuals for Visibility.* To further address the previously noted challenge of Desired Responders overlooking pertinent posts, we put forth another design idea that harnesses the Poster's close personal relationships with colleagues and peers, referred to as "Helpers". This approach aims to amplify the visibility of posts (DG2, DG3, DG6). This idea is illustrated in Fig. 1, D. Our formative study provided insight into this idea: participants expressed an interest in supporting their community connections. They indicated that upon encountering a post from their friends to which they couldn't directly contribute, they often redistributed it via their own accounts or directed it to Potential Responders within their professional circles. Thus, by capitalizing on these strong social ties, we can engage friends willing to help route the post to the right audience. Within this framework, the identity of the poster remains confidential, shared exclusively with the Helpers.

## 5 LITWEETURE SYSTEM

In this section, we describe our system implementation and the application of our proposed meronymity framework in the context of public paper recommendations within academic communities.

### 5.1 Example User Scenario

Alice, a beginning graduate student engaged in AI research, realized the need to delve into literature pertaining to "Trust in AI" to enrich her project's foundation. Yet, she hesitated to engage directly with domain experts and encountered challenges in gleaning pertinent paper recommendations from her academic circles. To bridge this gap, Alice opted to join LiTweeture, which offers a platform to connect with a broader spectrum of experts using meronymity, reducing her fear of social interaction.

Prior to submitting her inquiry, Alice added her advisor, Mark, as an Endorser (Fig. 2, A-1) to accentuate the credibility of her question, which he accepted. Subsequently, she articulated her question and composed a meronym by selecting four of her identity signals for public display of her identity: her junior status, her number of citations, Mark's publication to CHI, and Mark's number of citations. She desired a response from a renowned AI Expert, Rita, and chose to privately disclose to Rita her academic association with Mark (given his past affiliation with Rita's research lab), without revealing her identity. Additionally, she designated her colleague, Dave, as a Helper on this question and finally submitted her request to LiTweeture (Fig. 2, A-2).

On behalf of Alice, LiTweeture disseminated her query publicly across Twitter and Mastodon with her identity signals spanning two tweets in the same thread due to character limit, while simultaneously apprising Rita and Dave (Fig. 2, A-3) via email or DM on Twitter based on the provided contact for each person. On receiving an email invitation about an inquiry from an individual mentored by her former collaborator, Mark, Rita felt compelled to respond with her suggestions (Fig. 2, A-4). Concurrently, Dave was prompted via a direct message on Twitter to amplify Alice's inquiry. He subsequently quoted the post (Fig. 2, A-5), indicating that the asker is a close friend of his. Matt, a peer of Dave's who works in AI, felt inclined to respond when he encountered Dave's retweet of the original post as he was scrolling through his Twitter. Yet, harboring his own reservations about revealing his identity, Matt leveraged LiTweeture to share his recommendation under a meronym as well (Fig. 2, A-6). Upon Alice's approval of the meronymous contribution, LiTweeture posts the response to the conversation thread on Twitter and Mastodon (Fig. 2, A-7). An illustration of the final Twitter thread of this conversation is shown in Fig. 2, B.

### 5.2 LiTweeture Features

LiTweeture leverages the identity verification mechanism offered by the academic literature platform Semantic Scholar.[7] During the registration process, LiTweeture directs the user to log into this platform and claim their author profile page. Semantic Scholar employs a dedicated quality assurance team responsible for authenticating author profile claims, to ensure that the user who claimed a profile is its true owner. They also oversee the consistent updating of publication details of authors, including metadata such as publication dates, venues, and co-authors. Consequently, LiTweeture can reliably confirm the user's ownership of the claimed author profile and collect a verified record of their academic activity and publication history. Nonetheless, individuals lacking a publication record, such as emerging student researchers, will not possess an author profile on this platform to claim. Under such circumstances, access to LiTweeture's functionalities is contingent upon the addition of an Endorser, one who holds an author profile page and consents to the endorsement request.

To complete the registration process, users are asked to provide some personal details including their full name and Twitter handle. LiTweeture then verifies users' ownership of the Twitter account by sending a verification link via direct message to the provided Twitter handle.

*5.2.1 Endorsement.* The Endorser functionality enables users to bolster the credibility of their public identity and, in instances where a user's author profile page is lacking, ensures access to LiTweeture's functionalities. Thus, we can achieve identity verification either directly or through an affiliated Endorser. Even those with authenticated author profiles may still want to add an Endorser to augment their identity representation. The dynamic between the Endorser and the Endorsee operates on a one-to-many basis. While each user is constrained to incorporating a single Endorser at any given moment, an individual with a claimed author profile page can endorse several others.

---

[7] https://www.semanticscholar.org

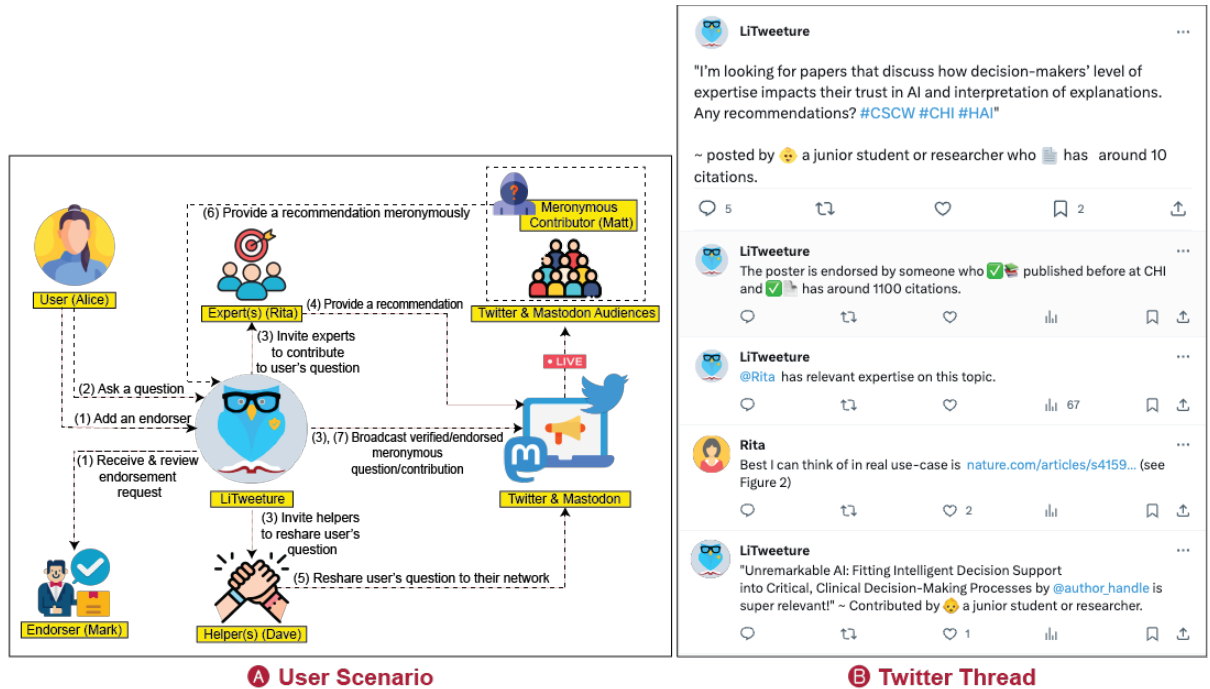

Figure 2: Example User Scenario on Asking For Paper Recommendations Using LiTweeture. Part A illustrates the steps of the user scenario. Alice requests Mark as her Endorser who then approves the request (A-1). Alice composes a question with a meronym, enlists Expert Rita with a personalized meronym, and enlists Helper Dave (A-2). Alice's question is posted with a meronym on Twitter & Mastodon on LiTweeture's accounts (A-3), Rita is privately messaged with a meronym (A-3), and Dave is privately messaged (A-3). Rita then responds publicly on Twitter (A-4). Dave then reshares Alice's question to his Twitter network (A-5). Matt, a potential responder, encounters Alice's question and provides a recommendation meronymously (A-6). After Alice's moderation, his meronymous contribution is posted publicly to Twitter & Mastodon (A-7). Part B illustrates the resulting Twitter thread from this user scenario.

Users can add an Endorser by providing the URL to the Endorser's author profile page and their contact email or Twitter handle (Appendix A, A). Endorsers are then notified of endorsement requests through email or Twitter Direct Messages. They are then prompted to describe their association with the Endorsee (Appendix A, B). LiTweeture offers four different relationship categories, each indicative of varying degrees of affiliation between the Endorser and the Endorsee: Advisor, Collaborator, Labmate, and Fellow Researcher. Furthermore, Endorsers can permit or prohibit Endorsees from revealing their name, publicly or to Desired Responders, by toggling a specific checkbox. Prior to accepting the request, Endorsers are presented with an interactive Tweet presentation illustrating what personal signals they are availing to the Endorsee, to make clear what information they are opting to share.

Both parties, the Endorser and the Endorsee, retain the discretion to amend or revoke the endorsement as they see fit. Upon receiving an Endorser, users are accorded access to the academic identity indicators affiliated with their chosen Endorser, and they are allowed to deploy these signals either in public queries or private messages to Desired Responders. The nuances of these identity signals will be elucidated in Section 5.2.2.

5.2.2 *Asking a Question.* Users with verified accounts are granted access to LiTweeture's asking functionality. Initially, users craft their inquiry. As this process unfolds, an interactive tweet visualization dynamically demonstrates the appearance of the prospective tweet or thread. This visualization accounts for Twitter's character constraints, seamlessly introducing additional tweets to a thread should the content exceed a singular tweet's limits. Subsequently, the user navigates through three pivotal stages: constructing their meronym, selecting Experts, and inviting Helpers.

**Constructing a Meronym.** Upon finalizing their inquiry, users are presented with an array of academic identity descriptors from which they can select to compose a meronym (Appendix B, A). Although our meronymity model encompasses arbitrary descriptions, we chose to focus on verifiable academic identity descriptors which seemed relevant to this particular meronymous application. These self-describing academic identity descriptors encompass elements such as the user's full name and Twitter handle, academic affiliations, citation and publication metrics, academic seniority, prior publication venues, and co-authorship networks. Examples of each category can be found in Table 1. While the user's name and Twitter handle are derived from the registration data, the subsequent signals are calculated using the publication history acquired from the Semantic Scholar public API. For users with an Endorser, they also have the privilege of accessing the Endorser's self-descriptive signals from this comprehensive list, which are procured in a similar manner. Additionally, users have the opportunity to disclose the nature of their relationship with the Endorser.

During our iterative process in curating and refining this identity signal list, we consulted four senior scholars for insights regarding LiTweeture and the incentives behind certain signals that would foster their engagement. One Associate Professor encapsulated the overarching sentiments by stating, *"If it's a student like, they are kind of guilting the person into saying, you should help this student because they look up to you. If it is someone that follows you or that cited you or something like that, it feels like [they're] my fan in some way. I should help them. If it's somebody like [reputable expert's name], for example, that I look up to him and he is more senior than me...It would be a way to help somebody who has visibility in the community."* Based on this feedback, we proceeded with our curated signals (Table 1) believing that they effectively capture the diverse facets of an academic's identity, fostering richer engagement.

LiTweeture empowers users to peruse and pick from this comprehensive set of identity signals to best represent their academic persona with any combination of signals. As users toggle between these signals, the tweet's content dynamically adapts, reflecting these changes in the interactive tweet preview. Once users are content with their identity representation, they can transition to the subsequent phase.

**Experts (also known as Desired Responders).** Users have the capability to extend invitations to multiple Desired Responders (titled simply Experts in the system) concerning their queries. To facilitate this, users enter the Expert's author profile page URL and provide a contact medium, either a Twitter handle or email (Appendix B, B). As a safeguard against potential misuse or misrepresentation, we employed heuristics to check for name similarities between the provided author profile URL and the contact details. While this method has limitations, we chose not to dwell on it extensively given the robust identity verification integrated within LiTweeture and the scope of our study.

The Expert's data is harnessed to amplify relational identity signals between the Poster and the invited responder, affording the user the opportunity to tailor their identity for individual responders. For every Expert added, a preview of the message, whether Twitter DM or email (depending on the provided contact), is shown alongside their details. Concurrently, a dropdown list (Appendix B, B), with the self-descriptive signals of both the user and their Endorser (if applicable), is provided.

Relational identity signals are derived by analyzing the publication history and Twitter interconnections of both the user and the Expert and added to the dropdown list. Such signals encompass bidirectional relationships of citation, co-authorship, and follows between the Poster and the Expert (Table 1). If the Poster has an Endorser, the system also formulates relational signals between the Endorser and the Expert, incorporating them with other signals. This dynamic mirrors scenarios like an advisor introducing their student to senior colleagues at conferences. This sentiment was reaffirmed by an Associate Professor we interviewed, who described that if a student tells them *"I'm a student of John, let's say I don't know the student, but I'm like, well John is somebody I know then it's less likely that I will ignore that email."* For every Expert, users retain the discretion to privately disclose specific identity signals to them. Additionally, users can also opt to reveal their name to them in contexts where they are comfortable. Finally, we note that messages are sent by LiTweeture not the Poster.

**Helpers.** In the final step of the asking process, users can add Helpers by simply inputting their contact details (Appendix B, C). The invitation, whether dispatched via email or Twitter DM to the Helper, transparently discloses the identity of the Poster. Such disclosure aims to foster motivation in the Helper, encouraging them to support their friend and potentially amplify the query by resharing the post. Consistent with the previous stages, a preview of the message is shown next to each listed Helper.

**Spam Control.** To safeguard against potential misuse of LiTweeture for spamming, we imposed specific constraints. Users can add no more than five individuals per category ('Expert' or 'Helpers') for each request. We have also capped the daily request count that an Expert can receive, ensuring that prominent people aren't inundated with excessive requests. Furthermore, users are restricted to sending out a maximum of three requests daily.

**Cross-posting and Various Modes of Communication.** Upon submission of the request, LiTweeture cross-posts the tweet or thread of the question to dedicated LiTweeture accounts on Twitter (Appendix B, D) and Mastodon, and notifies Experts and Helpers of the request. Our initial version of LiTweeture worked based off Twitter only, however, by the beginning of our field deployment study, the migration of academics from Twitter to Mastodon was significantly rising. Therefore, we incorporated cross-posting to Mastodon besides Twitter in an attempt to address this challenge. Besides cross-posting, we incorporated two modes of privately contacting others through LiTweeture: email or Twitter DMs. This enables communication with scholars on a variety of platforms, thus expanding outreach to a broader set of people.

**Public Enlisting of Experts.** In addition to private requests for answers from Experts, LiTweeture also enables the ability to @-mention Experts publicly in the Twitter thread to further grab their attention. From our discussions with senior scholars, some felt that directly asking senior experts publicly for help might motivate them to respond, as this may apply a sense of social obligation to help and also the answer may be broadly useful to the community. However, others expressed concerns about feeling pressured, finding direct requests for answers uncomfortable and akin to being "put on the spot." Thus, we phrased the public mention of Experts to be appreciative and inviting ("@expert_handle is knowledgeable on the topic of this question"), so as to be seen as a recognition of their relevant expertise rather than a direct request.

| Category | Involved Parties | Sub-categories | Some Examples |
|---|---|---|---|
| Self-Descriptive Signals | * Poster<br>* Endorser | Name & Twitter Handle | * Posted by Harry Potter, @harry_potter<br>* The poster is endorsed by their advisor, Severus Snape, @potion_master |
| | | Academic Affiliations | * Posted by someone who worked or is working at MIT<br>* The poster is endorsed by someone who worked or is working at MIT |
| | | Citation & Publication Metrics | * Posted by someone who has around 25 citations<br>* Posted by someone who has 1 to 5 publications<br>* The poster is endorsed by someone who has 500 citations and more than 15 publications |
| | | Academic Seniority | * Posted by a junior student or researcher<br>* The poster is endorsed by a senior professor or researcher |
| | | Publication Venues | * Posted by someone who published before at CHI<br>* Posted by someone who has 5 publications at CHI<br>* Posted by someone who published at CHI 2023<br>* The poster is endorsed by someone who has 20 publications at CHI |
| | | Co-authorship | * Posted by someone who co-authored with Hermione Granger<br>* The poster is endorsed by their advisor who is someone that co-authored with Albus Dumbledore |
| Relational Signals | * Poster - Desired Responder<br>* Endorser - Desired Responder | Citation | * The poster is someone who cited you before or someone you cited before<br>* The poster is endorsed by someone who cited you before or you cited before |
| | | Co-authorship | * The poster is someone you worked with before<br>* The poster is endorsed by someone you worked with before |
| | | Follow | * The poster is someone who you follow or who follows you on Twitter<br>* The poster is endorsed by someone who you follow or who follows you on Twitter |

Table 1: Classification and Examples of Identity Signals in LiTweeture. This table presents the types of identity signals that LiTweeture makes available for users, categorized into self-descriptive and relational signals. It details the involved parties and provides sub-categories for each type of signal, along with examples.

**Adding Experts and Helpers to Requests After a Post is Made.** LiTweeture allows users to add more Experts and Helpers to their requests at a later time after publishing their request using an identical interface to the one previously explained. This is to address the situation when the Poster learns about more Experts as they get more recommendations from the community.

*5.2.3 Answering a Question.* LiTweeture provides a way to answer questions using meronymity to motivate shy contributors to step in as well (Fig. 3). The functionality, interface, identity signals computation and implementation details of this feature are developed exactly as the question-asking experience described previously. An additional modification here is identifying if the contributor is the original Poster of the question and expressing this within their contribution. This is implemented to enable Posters to follow up on the conversation if they desire, *e.g.,* to thank a responder or ask a follow-up question.

*5.2.4 Moderation of Meronymous Answers.* Meronymous answers are not broadcasted instantaneously to Twitter and Mastodon. Instead, they undergo moderation by the original Poster. If they choose to approve the contribution, the answer gets broadcasted, otherwise, it is discarded. This feature ensures the elimination of spamming or trolling of Posters publicly.

*5.2.5 Aggregated Questions View.* Each Poster has an aggregated view that lists all their previously asked questions on LiTweeture for the ease of accessing requests.

## 5.3 System Implementation

The system's client is implemented using React and TypeScript, which connects to our server implemented using Flask. The server interfaces with a PostgreSQL database. Additionally, different functionalities are provided by using a number of APIs. The Semantic Scholar public API is used to obtain publication history data about users. The Twitter API is used for collecting network information about users for relational signals computation as well as for broadcasting requests on behalf of users. The Mastodon API is used to cross-post tweets to Mastodon.

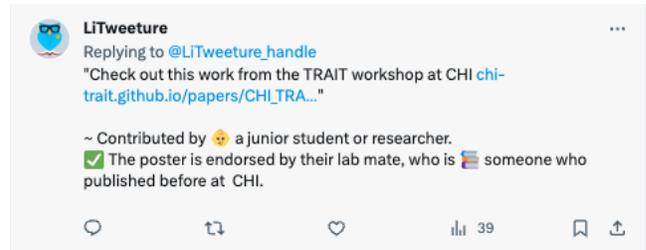

Figure 3: An Illustrative Tweet of a Meronymous Contribution Posted Through LiTweeture. The tweet is posted from LiTweeture's Twitter account on behalf of the original contributor. The tweet incorporates the contribution message with a paper recommendation and the meronym composed by the contributor.

## 6 FIELD DEPLOYMENT

To collect empirical evidence on how the proposed meronymous sharing design affordances impact and benefit the question asking and answering experience among academic communities, we conducted a one-month field study with 13 junior scholars. To facilitate the study, we also gathered the consent of a number of experts who were willing to be available to answer questions and added them to LiTweeture's interface as suggestions for participants to add them as Experts if they found them relevant. Our study received Internal Review Board approval.

## 6.1 Research Questions

In this study, our goal was to better understand and uncover insights regarding participants' engagement with and perceptions of the core design ideas of LiTweeture for help-seeking and contrast with their existing strategies. We also sought to deepen our understanding of participants' identity representation, their perception thereof, and the ramifications of meronymity throughout the experience. Our research questions encompassed:

- To what extent do our design ideas incorporated in LiTweeture shape participants' perceptions and experiences when asking questions and receiving answers, and when viewing others' questions and providing answers?
- In what manner do participants represent and interpret identities?
- What influence does meronymity exert on participants' public interactions and experiences?

## 6.2 Procedure

We conducted a field study in which participants were instructed to complete four tasks related to asking and providing research paper recommendations over a one-month duration. By having a field deployment over a longer period, we can get a more realistic understanding of the impact of meronymity on the help-seeking experiences of scholars.

Our field study was also a within-subjects experiment, encompassing two weeks using LiTweeture to complete two tasks and two weeks using a baseline condition to complete two tasks, with order counterbalanced. This enables us to contrast the use of LiTweeture with the status quo of asking and answering questions, where in the baseline condition, participants have the autonomy to select their preferred methods of inquiry, encompassing any tool and method, inclusive of both private and public interactions. We also ask all participants for four paper recommendation questions up front so that questions are not tailored to specific conditions, enabling us to more easily compare analogous tasks across the two conditions.

*6.2.1 Conditions & Task Assignment.* Upon registration, participants were allocated to a condition, delineating their study trajectory, on-boarding process, and sequence of tasks. Participants were then prompted to arrange an on-boarding session. All study meetings and interviews were conducted over Zoom. A diagram illustrating the full structure of the study is shown in Fig. 4.

During the on-boarding process, participants completed an initial survey where they were instructed to provide four questions pertaining to scientific literature in domains of their interest. Each question they provided then became a question they would ask in each of the four tasks.

Participants were asked to accomplish two tasks employing LiTweeture (System condition) and two tasks using any alternative method of their preference (Baseline condition). We counterbalanced participants to complete either the two tasks in the System condition first or the two tasks in the Baseline condition first, to mitigate potential order effects. The two tasks for each condition were to be executed consecutively, each extending over a week, and then participants transitioned to the other condition to complete the remaining two tasks in the next two weeks. The four questions participants provided were assigned in random order, i.e., participants could not pick which questions to ask in which condition or in which week within each condition.

Initially, we enlisted 16 participants to achieve a balanced task assignment in both conditions. However, three individuals withdrew post-registration, and an additional three withdrew during the course of the study subsequent to completing the initial condition. This resulted in a slight disequilibrium in condition assignments, such that 4 did System-Baseline, 6 did Baseline-System, 1 person did just System, and 2 people did just Baseline.

*6.2.2 On-boarding & Tutorial Meetings.* Both conditions had an on-boarding session for 10–15 minutes. Participants were prompted to first complete a pre-study survey. In addition to providing four questions, they gave details about their engagement in content sharing on Twitter, their experiences asking for paper recommendations, and provided demographic information.

For those in the System condition initially, the interviewer had them create an account on LiTweeture and provided the option to invite an Endorser, if so desired. A second meeting was scheduled a few days later (Tutorial session) spanning 30 minutes to walk them through LiTweeture and delegate the initial pair of tasks. The necessity for separating the on-boarding and tutorial sessions arose due to LiTweeture needing claimed Semantic Scholar author profiles, which can take a few days to validate. Additionally, this interval ensured Endorsers had time to respond to requests. During the tutorial session, we demonstrated the features of LiTweeture to the participants. Participants then completed a practice task while screen sharing where they asked a question on LiTweeture employing a meronym and inviting a minimum of two Experts and two Helpers to the request. Additionally, participants were shown how to answer requests on LiTweeture meronymously.

Participants in the Baseline condition initially also were asked to make an account on LiTweeture and claim their Semantic Scholar author profile. However, they did not have access to the full suite of LiTweeture functionalities. Instead, they were presented with a Baseline condition interface that simply enabled them to document their completed control tasks via a form integrated into the page. Before the start of their System condition, they then went through the Tutorial session mentioned above.

*6.2.3 Tasks.* After on-boarding, participants were emailed specific instructions on completing the initial pair of tasks. In the System condition, each weekly task comprised two sub-tasks: asking the assigned question and answering any question of choice that had been posted by another participant publicly on the LiTweeture Twitter account. Participants were asked to invite a minimum of three Experts and three Helpers to their question. Incorporating prevalent hashtags, such as "#CHI2023", was recommended for broader visibility. Participants could select their identity representations for both sub-tasks and could choose to answer using LiTweeture or their personal Twitter account in the latter. In the Baseline condition, participants chose whatever method to ask their question. They then documented the message, method, and number of individuals contacted.

In both conditions, the asking sub-task needed completion within the week's initial two days, allowing time to accrue responses. The answering could be done at any point during the week. To conclude the task, participants completed a reflection survey about both sub-tasks within the last two days of the week. The reflection survey used 7-point Likert scale questions to evaluate factors such as the novelty of recommendations they received. The survey also asked for a description of their asking experience. In the System condition, participants further elaborated on: (1) their identity representation and sentiments regarding LiTweeture's features; and (2) their selected question to answer, the portrayal of their identity in responses, the rationale behind such portrayal, and their perception of the asker's identity. Participants received bi-weekly reminder emails for each task.

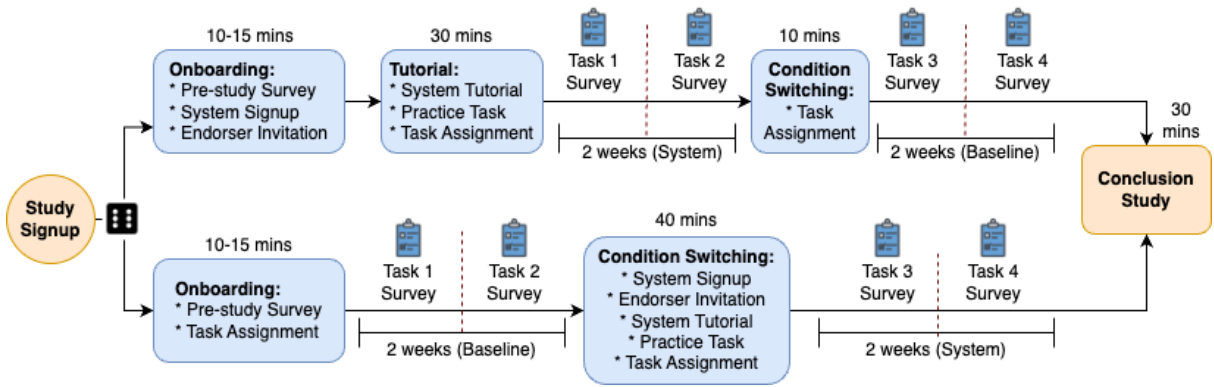

Figure 4: Overview of the Field Study Design. This diagram outlines the phases of the field study, detailing the on-boarding process, tutorial, task assignments, condition switching, and the concluding study components. It also indicates the duration of each stage and the overall study timeline, with participants being randomly assigned to study conditions upon signup.

*6.2.4 Post-Study Interview.* After all tasks were completed, participants were invited to a 30-minute interview session. During this interview, participants were prompted to discuss their experiences in both conditions. Specifically, participants were asked about their thought processes and sentiments concerning asking a question, viewing others' questions, and receiving and providing answers. Furthermore, participants were encouraged to delve into their perceptions and conceptions regarding identity, encompassing both self and others.

### 6.3 Participants

Our recruitment strategies included word-of-mouth, mailing lists, HCI groups on Meta, and advertising on personal Twitter and Meta networks of co-authors and colleagues. We sought participants who met these criteria: active researchers, past or affiliated with paper publications, Twitter users during the study, and adults aged 18 or above. Participants were compensated $20/hour (USD) on PayPal at the end of the study.

We enrolled 13 participants (8 female, 5 male) with diverse academic backgrounds, including 12 PhD Students and 1 Post-doc. Participants' ages ranged from 18–44, with 6 aged 18–24, 6 aged 25–34, and 1 aged 35–44. Participants were primarily from the USA (11), with others from South Korea and Denmark. They had various academic specializations, including 11 in HCI, 1 in Computational Biology, and 1 in Formal Methods. Academic experience ranged from 1 to 5 years (median: 2), with 0–17 published papers (median: 5). Participants typically sought paper recommendations with varying frequency (3 rarely, 8 sometimes, 2 often) and demonstrated diverse Twitter activity (1 never, 3 rarely, 5 sometimes, 4 often).

### 6.4 Data Collection & Analysis

Task details were collected by logging participants' actions and recording their entries on LiTweeture. Out of the 13 participants, 10 completed all tasks in both conditions, with one participant completing an extra System task. Two participants completed tasks only in the Baseline condition, and one participant completed tasks only in the System condition. This resulted in 23 System tasks (23 asking sub-tasks & 22 answering sub-tasks, as one participant didn't complete their final answering sub-task), and 24 Baseline tasks, each with an associated reflection survey.

In the System condition, participants reached out to three Experts and three Helpers per task using LiTweeture, inviting a total of 69 experts and 69 helpers across all tasks. Notably, each participant in the System condition received at least one paper recommendation response during either of their tasks.

In Baseline tasks, participants employed diverse methods to request paper recommendations from experts. Predominantly, One-On-One digital messaging was utilized (N = 11 out of 24 tasks), followed by Group digital messaging (N = 5 out of 24 tasks), and Verbal in-person or video-conferencing communication (N = 5 out of 24 tasks). Additionally, Public communication on social media, forums, or blogs was employed (N = 4 out of 24 tasks). The vast majority of participants used a single mode of communication, with the exception of one participant who employed both one-on-one digital messaging and verbal in-person or video-conferencing communication within a single task. In aggregate, a total of 91 Experts were contacted across all Baseline tasks (N = 24). Two participants in the Baseline condition did not receive any recommendations in either of their Baseline tasks.

To ensure an equitable comparative analysis between the two conditions, all descriptive statistical comparisons between them exclusively encompassed the subset of 10 participants who successfully completed all tasks in both conditions, amounting to 20 tasks for each condition. We designate this group as the "Completion Group" and their tasks as "Completion tasks" for future reference.

Taking only the "Completion Tasks" in the System condition (N = 20), the cohort of 10 participants collectively invited 60 experts and 60 helpers to their questions. The paper recommendations received by this group ranged from zero to four ($\mu$ = 1.4, $\sigma$ = 1) per task. Similarly, in the "Completion Tasks" within the Baseline condition (N = 20), the same group of 10 participants interacted with 74 experts, ranging from one to six ($\mu$ = 3.7, $\sigma$ = 2.1) per task. The paper recommendations received in these tasks varied from zero to seven ($\mu$ = 1.8, $\sigma$ = 1.6) per task.

Free-form text responses from surveys and the post-study interview were analyzed using an inductive thematic approach. The primary author conducted multiple iterations of the analysis and engaged in discussions with co-authors to refine the code-book as necessary, ensuring consensus was achieved. We chose not to calculate inter-rater reliability because the codes were developed as part of a thematic analysis to yield concepts and themes [62]. In the subsequent sections, we provide participant quotations, identified by strings in the format "P-participant identifier."

## 6.5 Results

This section presents our findings from the field study. Initially, we highlight the influence of meronymity on various facets: confidence, contextualization and specificity, credibility and trust, and accountability of participants. Subsequently, we demonstrate participant concerns regarding the oversharing of personal information and certain identity signals. Following this, we showcase meronymity's impact on the direct asking of experts and the concern of participants about overburdening them. Lastly, we outline the collective perception and role of LiTweeture, alongside participants' experiences with it.

### 6.5.1 Meronymity.

**Meronymous interactions elevated participants' confidence and social validation, while delivering quality recommendations.** Meronymity provided participants with the benefits of anonymity. One prominent theme that emerged in the Baseline condition was *"social anxiety"* and *"fear of judgment and reputation concerns"* with openly seeking guidance, which corroborates the formative study findings. Most participants expressed apprehension about being perceived as *"lazy or not a good researcher"* or having their questions deemed irrelevant for a public space when asking for literature recommendations. For instance, P-4 articulated concerns about how asking certain questions might reflect on their expertise in the community: *"What's the community going to think... Isn't she supposed to maybe be one of the experts about this?"* This prevalent fear of seeking public assistance was counteracted in the System condition by the advantage of meronymity in overcoming social hurdles and in *"feeling more free to ask"*. As one participant aptly put it, *"while I felt a bit uncomfortable reaching out to the experts I did, it was really easy to do so anyways since my identity was hidden." (P-10)* LiTweeture's meronymous nature provided participants with a protective shield, as P-4 preferred going unanswered anonymously rather than publicly: *"So if I'm not going to get recommendations or many recommendations, I prefer that happens under an anonymous platform instead of reminding me that I asked something and nobody engaged with me".*

Participants expressed a deep sense of validation, excitement, and support through their engagement on LiTweeture and metrics like retweets and likes. P-6 shared their enthusiasm: *"seeing three retweets, I was like, oh my gosh. I felt really excited... It made me excited because people were like, oh, this is a good question. Why didn't I put my real name?"* Similarly, another participant felt validated when their tweet was quoted by an expert: *"I think regarding retweets, likes, and especially the quote retweet... it's nice to feel that I am asking relevant questions for the community."* The quality of recommendations and interactions on LiTweeture also stood out, with P-2 remarking on the impact of receiving an expert's answer: *"But [expert's name] answered my first question on Twitter as an expert, and that was really great to see his kind of answer."* Participants also found value in the social cues from interactions on LiTweeture, even without specific recommendations. They appreciated the social experience, as P-10 noted: *"...just seeing that engagement with those kinds of questions, I think it's a very positive thing for someone doing research."* The lack of many suggestions was sometimes seen as a confirmation of their work's novelty. P-4 reflected on this: *"I didn't get more suggestions because there is not that much more to suggest which is a nice confirmation on one hand that I'm working on something quite novel."*

**Meronymous identity facilitated contextualization and specificity in public interactions of participants.** Participants employed meronymous identities to strategically reveal aspects of their personal and professional selves, impacting the way they engaged and were perceived in the academic community. This selective revelation of identity aspects such as conference affiliations and seniority levels played a crucial role in obtaining specific and relevant responses to their queries. Participants often used LiTweeture to share partially anonymized identity signals, enhancing the credibility of their requests and encouraging tailored recommendations. Notably, some participants chose to highlight their publications in specific conferences relevant to their requests, aligning their queries with their academic expertise and community. P-3 shared the thought process behind this strategy, saying, *"I thought that specifying the name of the conference would be much more relatable."* Similarly, P-11 emphasized the importance of being associated with well-known conferences for credibility: *"CHI is very popular...that's why I think mentioning CHI would be a good way to ensure my credibility."* Influencing the quality of responses received was another goal. Participants like P-4 noted, *"when I go through all the descriptors for how I want to present myself, what I'm actually doing is thinking of what kind of recommendations I want to get."* This approach ranged from showcasing seniority within a domain to presenting oneself as a novice to garner foundational recommendations. Referencing endorsers or co-authors to bolster credibility was also a common strategy, despite the potential risk to anonymity.

On the responders' side, the identity signals from askers significantly informed and shaped their recommendations. Participants found these signals crucial in providing context and specificity to their responses. P-4, for instance, felt more inclined to suggest relevant papers when recognizing a scholar's publishing history: *"This* *person published before at CHI... It makes me feel that it's relevant to suggest CHI papers to this person."* In contrast, a lack of identity signals sometimes resulted in a decline in motivation and confidence in providing responses, as highlighted by P-6, who felt *"slightly less encouraging to answer"* without such cues. Identity signals extended beyond mere academic credentials; they often acted as relational bridges, fostering a sense of connection and motivating participants to engage more deeply with the queries. This was evident in cases where some participants expressed how an *"identity really inspired [them] to try to find a good reference"* and went beyond their expertise doing a *"mini-literature review"* to assist a junior student, driven by a shared academic affiliation or experience. While some participants maintained an unbiased stance regardless of the identity presented, the overall sentiment suggested that identity cues played a significant role in the engagement process.

**Meronymous identity facilitated credibility, connectivity, and motivation in public interactions of participants.** When incorporating endorsers' identity signals into their requests, participants felt more confident, as they perceived endorsers added credibility and validity to their questions, thus making their questions more convincing. Participants also believed that endorsers played a crucial role in enhancing the perception of reliability. As P-2 remarked, having an endorser signified *"reliability around this person."* and that the additional social signals helped in *"humanizing"* their requests, making them appear more genuine and personal. In addition, participants also felt that revealing specific signals about their endorsers, such as the number of citations and publications, could further boost their own credibility to be *"viewed as a serious [but] junior student researcher."* For instance, P-6 mentioned how they strategically enhanced their credibility, stating, *"the number of citations and the publications I added just to make it more valid like the person that I am endorsed by is a really serious researcher..."* Furthermore, participants perceived that endorsers increased the likelihood of obtaining responses. As one participant explained, *"having the endorser... makes my question valid, but also will hopefully make the results more valid"* (P-6). Additionally, endorsers were seen as a means to expand one's social reach, tapping into the endorser's broader network and reputation, as mentioned by P-12.

Participants appreciated the flexibility in balancing between revealing relevant information and maintaining anonymity. P-12 highlighted this balance, stating, *"I think [the way I presented my identity on LiTweeture] keeps a good balance between anonymity while also showing people that me and my endorser have the relevant experience in the field to prove to people that we are part of this community and we are asking questions that matter to us."* This strategic identity presentation was echoed by others, like P-4, who felt comfortable sharing co-authorship details, and P-5, who believed that disclosing conference affiliations didn't compromise their anonymity.

Participants also used private relational identity signals to establish connections with individual experts. Employing meronymity, they leveraged personal connections and the reputation of endorsers to increase the likelihood of receiving responses. P-2's approach of identifying experts with whom they had shared connections exemplified this strategy: *"if I could identify [an expert] that has personally collaborated with [someone I know]...they might be more likely to respond to it."* This tactic was also evident in cases where participants revealed their full identity in one-to-one communications with trusted experts to foster a more genuine connection.

**Meronymous identity was leveraged by participants when perceiving and signaling quality and trust.** Participants' perceptions of recommendations were significantly influenced by the meronyms of the respondents. Recommendations from recognized experts were highly valued, as P-8 expressed: *"I definitely took the [recommendation provided on one of their questions] more seriously...the author is relatively well known in the field."* In contrast, meronymous recommendations from junior researchers were often met with reduced trust, with P-8 observing, *"somebody who only has one to five publications...gave me a social cue that this person is relatively junior,"* impacting their credibility. Furthermore, the meronyms of respondents often forged a personal connection. P-10 felt a sense of relation receiving recommendations from peers in similar fields: *"It's nice to see the people who are in FAccT who published in CHI before are also interested in what I've been looking at."* Such connections sometimes spurred interest in further collaboration, as P-12 contemplated deeper engagements: *"I think it's interesting...this person might be someone I want to know personally."*

When providing recommendations, participants also aligned their identity signals with the confidence and quality of their recommendations. Full identity disclosure was used in some instances when participants were confident about their suggestions. Conversely, meronymity was common as a strategic tool to signal quality without revealing full identity. For instance, some participants used their status as a *"junior student researcher"* to lower expectations regarding their recommendation's quality, while others mentioned their publications at prestigious conferences to hint at their expertise and confidence in the recommendation. Participants employed both full identity and meronymity based on their confidence in their recommendations and the impact of social factors, using these methods to communicate the credibility and quality of their interactions.

**Meronymity encouraged a sense of accountability among participants.** The implementation of meronymity in LiTweeture fostered an environment where participants, while benefiting from the protective cloak of anonymity, still felt a sense of accountability. Contrary to the conventional belief that anonymity might lessen the quality of input due to reduced pressure, the semi-anonymous nature of LiTweeture encouraged participants to be more vigilant in their communication. As highlighted by P-5, *"Even though I know it's semi-anonymous...I would check the grammar multiple times,"* emphasizing the care taken to ensure their queries were well-phrased and not overly naive. The participant further elucidated that a fully anonymous platform might have led them to more casual phrasing without the use of domain-specific jargon, which showcased their expertise. A salient aspect that stood out in participants' feedback was the presence of institutional prestige, such as being associated with a reputable entity. This association brought along its own set of pressures. P-7 candidly shared, *"If I'm gonna share like I'm from [university name]...and have a very stupid typo, I feel bad,"* underlining the fear of misrepresentation or tarnishing the image of the affiliated institution. In essence, LiTweeture's meronymous design prompted users to strike a balance between leveraging the freedom of reduced judgment while simultaneously upholding a level of personal and institutional responsibility in their communications.

**Concerns over excessive information sharing were voiced by some participants.** Participants expressed concerns over revealing too much identity information, balancing the need for recognition with privacy and anonymity. P-10 captured this sentiment, stating, *"I also didn't want to show too much of my identity, which is why I didn't choose to make my institute or my university public or who I've collaborated with."* This cautious approach was common, with participants fearing unintended disclosures and breaching collaborators' privacy. P-13 explained their hesitation: *"I debated between putting in my collaborators... But I was a bit afraid of breaching their privacy because I didn't really talk to them about using their names."*

The reasons for this cautious behavior varied. P-5 was concerned about public perception and appearing intellectually superficial with *"less thought-out"* questions, saying, *"...I am just a little shy to just disclose too much information about me."* Participants also differentiated in their willingness to share information based on their relationship with experts, with P-11 revealing more to familiar contacts. When inviting Helpers, a similar sentiment was expressed by some participants who felt that private disclosure of identity to Helpers compromised their sense of anonymity.

We noticed over time, some participants reported growing comfort with disclosing more information about identity within LiTweeture. This suggests that initial unfamiliarity with LiTweeture may have led to concern, and with time and exposure, the disclosure of more information about identity became less concerning.

**Participants provided additional feedback on identity signals.** Most participants appreciated the extensive list of identity signals provided by LiTweeture. As P-2 commented, the system had a *"very exhaustive, nice list"* that was *"super relevant, super representative."* While the majority perceived signals positively, some participants worried about potential biases in some of the signals. For instance, P-7 questioned the reliance on metrics like publication and citation counts, noting that the metric might be biased and may not reflect a researcher's true worth or the importance of their questions. Additionally, some participants noted that using endorsers' identity signals sometimes provided redundant information or was not their primary focus. High-citation endorsers were also seen as *"intimidating"* by some, possibly deterring engagement.

Participants suggested additional identity signals for better representation. The inclusion of gender representation was suggested, with one participant emphasizing the potential value of incorporating gender in receiving help and *"starting conversations,"* especially in fields with for example *"less number of women"* and other gender minorities. Geographical identity was also seen as valuable, with one participant wanting to highlight their status as a *"researcher in Asia."* Participants recommended adding educational or professional background for enhanced relatability. Others desired more specific indicators of research specialization and subjective self-descriptions of knowledge levels, like one idea of having a *"very basic understanding about machine learning."* Additionally, indicating personal interests or goals, as P-4 suggested, could help provide a clearer context for the questions. Overall, while the current identity signals were largely well-received, participants saw room for improvement, calling for a more diverse set of signals to address concerns of bias and better represent their multifaceted identities.

*6.5.2 Experts and Helpers.*

**Meronymous direct asking boosted confidence and visibility of participants, thereby yielding novel recommendations.** The familiarity of participants with contacted experts considerably varied between the control and treatment conditions (Fig. 5). In the control condition, participants gravitated towards seeking guidance from familiar contacts. Notably, most Completion Group participants (N = 7 out of 10) exclusively consulted friends or those they shared strong relationships with. The reasons ranged from the comfort and ease of asking close contacts, as P-5 succinctly noted: *"I liked that I was quite comfortable texting my friend for the request,"* to reservations in extending their outreach to unfamiliar experts. One participant expressed hesitancy to approach reputable researchers: *"There are some experts I could ask, but they are researchers in industry that I am not comfortable emailing to ask about literature recommendations."*

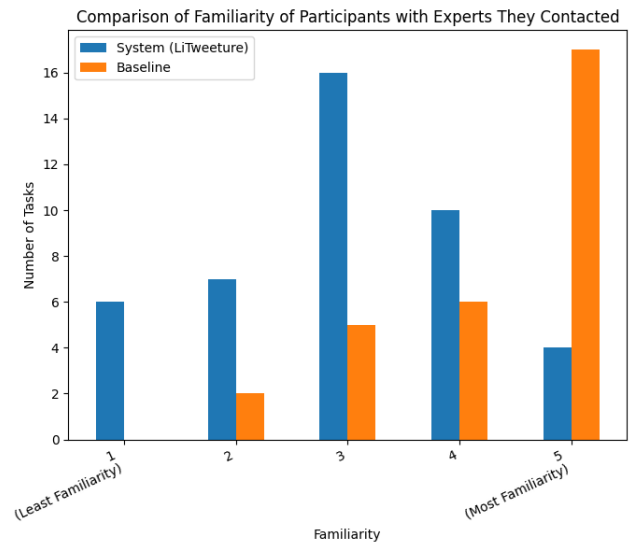

Figure 5: Comparison of the level of familiarity of participants with experts they contacted between Baseline (N = 20) and System tasks (N = 20). Participants reached out to more experts they have more distant relationships with in the System condition and none of the participants contacted experts they have not talked to before in the Baseline condition. Note that the y-axis denotes the number of instances each familiarity category applied among the 20 tasks per condition.

In contrast, LiTweeture appeared to lower these barriers and expand participants' horizons in terms of outreach. A majority of participants (8 out of 10 of the Completion Group) approached all their desired experts, signaling an increase in comfort. The system enabled users to engage with esteemed scholars they'd otherwise find daunting. P-10's reflection encapsulates this shift, *"I liked that I could invite experts that I knew of but might feel a bit intimidated by... I felt like this was a good opportunity to somehow connect with these experts."* Furthermore, LiTweeture was perceived by some participants as a conducive platform to revitalize slight interactions, turning them into meaningful exchanges. As P-7 indicated, the system felt like *"an interesting way to experiment"* with reaching out to acquaintances from conferences, turning occasional interactions into avenues for academic discussions without the awkwardness that direct messaging or emailing might entail. Participants also appreciated the system's ability to amplify the visibility of their questions through the support of friends or colleagues. P-2 highlighted this benefit: *"having friends that would also help forward or boost your posts, that was really helpful too."*

As a result, participants were able to tap into a broader expert community through LiTweeture which yielded more novel recommendations. In the Baseline condition, a significant number of participants reported receiving recommendations that, while relevant and useful, were not novel to them. This familiarity seemed to be a by-product of frequent interactions within close-knit academic circles. In the System condition, many participants acknowledged LiTweeture's role in facilitating broader connections. For instance, P-4 highlighted the dual advantage of reaching a broader audience while retaining a semblance of anonymity: *"But I would also love to get access to a wider audience outside my direct network. And yeah, I like the idea of wearing a little bit of a mask..."* Completion Group Participants (N = 10) reported, on average, a higher rating (from 1–7) of novelty of recommendations in the System condition ($\mu = 5.33$, $\sigma = 1.26$) compared to the Baseline condition ($\mu = 4.6$, $\sigma = 1.56$). One participant noted, *"Those were papers that I didn't necessarily know of before, so they were brand new and through the papers I was also able to get more papers"* (P-13).

**Participants had concerns about overburdening experts.** In both conditions, participants voiced genuine apprehensions about overburdening established experts and helpers with inquiries, fearing the social implications of doing so. Participants pointed out, experts are often preoccupied with their own academic endeavors and teaching commitments, making them wary of adding to experts' load. This issue was further reinforced by other participants when reflecting on LiTweeture, *"I'm worried that they are busy and I imagine that they receive a lot of such kinds of requests everyday"* (P-5). The concern was not merely about being disregarded due to the expert's assumed inundation with similar requests, but also about the perceived expectation in academia to thoroughly research a topic before reaching out as reflected by one of the participants: *"..I feel a little guilty that I didn't do any homework..."* Similarly, some participants expressed reservations about inviting Helpers due to concerns of intruding on peers' time and attention. One feature of LiTweeture that helped to mitigate some of these concerns and that garnered appreciation from multiple participants was providing a set of experts who had willingly given their consent to be approached. Participants emphasized the importance of not wanting experts to contribute *"free labor"* and appreciated that the platform was populated with experts who had agreed to be a part of the LiTweeture community.

*6.5.3 Overall LiTweeture Experience.* Our findings from the study underscore the positive sentiments participants held regarding LiTweeture's platform in comparison to alternative methods. Notably, the system was acknowledged for its ability to lower the barriers traditionally associated with online public interactions, especially in the context of content sharing and seeking assistance. For instance, P-10 shared, *"I liked that it was easy to post things on Twitter, which I would probably never do on my own personal account."*

Participants expressed their interest in using LiTweeture as *"a great way to source papers"* in the future. One participant, who publicly asked a question from their personal account on Mastodon, remarked, *"I know I could've used LiTweeture. And I wish I did because it probably would've gotten me better results."* LiTweeture also acted as a resource for other members of the community which was indicated by one participant stating *"The questions I saw were very interesting, some of the questions are questions that I had but I just didn't think to ask them. And going through the questions, I just thought I might come back to them. These contributions would also be an archive of someone who's trying to explore that certain topic"* (P-13).

Although participants expressed their interest in using LiTweeture in the future, the introduction of LiTweeture into the academic space appears to serve as a complimentary tool rather than a replacement of traditional methods for seeking scholarly advice. The core distinction lies in the modality of communication. LiTweeture's strength lies in asynchronous communication, opening doors for junior scholars to pose broader questions without the intimidation factor. On the other hand, participants highlighted an inherent limitation: *"you can't go into DMs and talk about the topic a bit more."* The asynchronous nature of LiTweeture, while advantageous in alleviating initial barriers, does come at the cost of real-time, in-depth conversations which direct messaging facilitates. Incorporating meronymous DMing to LiTweeture could address this limitation. Participants noted that there was still a preference to directly *"reach out to colleagues"* when there was an immediate and specific requirement, suggesting that direct interactions still held weight in situations demanding immediate resolutions.

## 6.6 Summary

Our results underscore the power of meronymous communication on digital platforms, especially in communities with pronounced hierarchical structures. Some of our key findings include:

- Meronymity, relative to full identification, increased junior researchers' comfort in asking a wider range of experts for recommendations publicly.
- Strategically revealing aspects of one's identity provided important context that made askers appear credible and allowed answerers to tailor recommendations.
- The ability to partially-anonymously interact with a broader community was exciting and motivated participation.
- There remain opportunities to grow engagement on LiTweeture and refine identity signaling capabilities, but overall reactions were positive. The system demonstrated the potential benefits of designing for partially anonymous communication tied to real identities.

# 7 DISCUSSION AND FUTURE WORK

Our work contributes to resolving the tension between online safety and freedom of expression by introducing Meronymity, a novel design paradigm that allows individuals to reveal specific aspects about their identities in effort to address the social barriers to public participation faced by them in online communities like academia [11, 29, 34], which could stunt their development and narrow the diversity of public dialogue [11, 29, 34]. Although anonymity could be one solution, it often fails to generate tailored responses and tends to receive less engagement [31, 56]. On the other hand, persistent pseudonyms enable the accrual of reputation, relationships, and accountability over time but constrain flexibility in identity presentation and audience management [18, 49]. This limitation forces rigid trade-offs between long-term privacy and short-term contextual needs.

Our formative study with 20 scholars confirms the existence of public participation barriers in academia and uncovers a preference for partial anonymity that balances privacy and credibility. We propose a meronymity model, and develop and deploy a system, LiTweeture, incorporating it to address this need for flexible, context-aware identity control. LiTweeture serves as a tangible prototype of the meronymity design paradigm in action as it offers a realistic environment for exploring this innovative approach to identity representation in online conversations and help-seeking scenarios.

Our results reflect a positive impact of meronymity on public engagement while protecting privacy. This is achieved as meronymity combines the benefits of full identity disclosure (contextualization, specificity, credibility, connectivity, motivation, trust, quality and accountability) through disclosing meronyms and endorsements, and full anonymity (confidence and social validation) through hiding parts of users' identities. Our deployment revealed several advantages of this approach: (1) Participants tailored their identity disclosure for each interaction, enhancing response quality, (2) Selected academic credentials established credibility for junior members and allowed for customized answers—flexibilities unavailable in anonymous and pseudonymous systems, (3) The platform fostered a sense of validation, excitement, and community, enriching the social experience beyond typical one-on-one exchanges, and (4) Meronymity reduced fear of judgment; thereby encouraging inquires to experts and eliciting novel recommendations.

The overall sentiment was positive, affirming meronymity's potential to transform public questioning barriers for junior members. However, some concerns were voiced by participants around risks of oversharing identity aspects. In the following sections, we further reflect on the potentials and limitations of our system in the academic setting. We also discuss the design and real world implications of deploying meronymity in a general-purpose social platform.

## 7.1 Interplay of Meronymity with Community Building and Social Capital

*7.1.1 Supporting Meronymous Identities vs. Fostering More Supportive Communities.* One fundamental question around the motivation of this work is why partial anonymity is needed in the first place, and whether we should instead foster more welcoming online communities where *"newbies"* do not fear engaging with the larger community. While we agree that this is important, our results suggest that partial anonymity is valuable regardless. Arguably, almost all communities have some element of social hierarchy or status signals that can become a source of pressure. Our interviews with faculty members suggest that social pressure occurs at levels beyond junior researchers. One assistant professor said that *"[I care] about my public persona and how I want to be perceived, especially by my more superior peers."* Further, even in communities where old-timers are always friendly and supportive to newcomers, some newcomers may still feel the tension to engage. For example, one participant from our user study pointed their *"imposter syndrome"*. This suggests that newcomers sometimes experience fear of judgement regardless of how old-timers perceive them or their questions.

Additionally, in public spaces, such as Twitter, the boundaries of different communities can be porous. Even if one is in a supportive community, there is still risk that other users could see one's post and potentially be harassing. Relatedly, we found evidence that LiTweeture can potentially foster better community and discussion in public spaces such as Twitter because it provides a sense of belonging and a dedicated channel for seeking help from senior members that currently does not exist on Twitter.

*7.1.2 Enabling Accumulation of Social Capital with Meronymity.* Meronymity, while offering identity protection, can impede the accumulation of reputation and to some extent social capital that are often facilitated by full identity disclosure. To address this, we suggest a hybrid approach that blends meronymity with pseudonymity. This combination allows users to build and maintain a consistent presence, fostering recognition and trust over time, while still safeguarding their actual identities. Pseudonyms provide an alternative mechanism for obscuring identity. In contrast to a meronym which generally describes a *class* of individuals satisfying the description, a pseudonym is bound to a *unique* individual, although which is unknown. This has the benefit of permitting that pseudonym itself to accumulate reputation and/or social capital through interactions with others. A meronym cannot similarly accumulate reputation, as many individuals may be posting under the same meronym. On the other hand, a pseudonym can be attributed to a negative reputation based on past interactions, making it challenging to disclose only context-specific identity information in a given discussion [49]. Previous research has shown that users create single-use throwaway accounts to get around this challenge [49]. In other occasions, a (pure) pseudonym provides no (even partial) description of its user, meaning that it suffers from the same problems as anonymous communication: reduced motivation to engage, and lack of understanding of the poster that could help provide context for the engagement. Thus, the two types of nym have different benefits. Of course, they can be combined; one could associate a meronym with a pseudonym, creating a context-specific identity that would present useful information for initial engagement with other individuals but could also accumulate reputation and social capital over the course of repeated engagements. This would also address the challenge of attribution to negative reputations by enabling users to easily joggle between their various online personas without the burden of creating throwaway pseudonyms.

## 7.2 Risks of Meronymity and Potential Mitigations

*7.2.1 Calibrating Social Pressure To Avoid Overburdening Users.* Social pressure is not uniformly bad. Social pressure can beneficially motivate people behaving properly towards each other. Deciding whether social pressure is too high—discouraging participation in community conversations—or too low—permitting anti-social behavior—requires consideration of the context and may be debated between different parties. However, our interviews strongly suggest that social pressure in the specific setting of academic discussion is too high, with especially junior members feeling reluctant to participate in public, even when that participation would be welcomed by others including senior members. Therefore, we have explored the introduction of ideas that would reduce social pressure, and our results showed that these ideas were successful.

It is possible that these techniques could be *too* successful. Meronymity might encourage people to burden others with questions that they could answer themselves with little effort. In a related vein, as we saw in our results, meronymity allows people to remain invisible while consuming help from the community, which may let them "freeload" without social penalty, getting help but giving none. Signs of going too far in reducing social pressure might be when community members feel that too many "inane" or "lazy" or "unhelpful" posts are being made meronymously or when community members are getting far too many requests in general. Were this problem to arise, we could consider mechanisms to bring social pressure back up, *e.g.,* by discouraging senior members from being quite so free with their endorsements or blocking or ignoring endorsers who are too free with their endorsements.

Experts could also be given affordances to manage their availability and engagement. This includes options to control the volume of incoming requests, temporarily opt out during busy periods, or delegate questions to other qualified researchers or students. This approach respects experts' time and commitments, while ensuring continuity in assistance for askers.

Additionally, to enhance efficiency and reduce redundancy, the system could be designed to prompt users with similar questions that have already been addressed by their chosen experts within the LiTweeture framework. This feature would help streamline information exchange, ensuring that efforts are not duplicated and that users can readily access existing knowledge.

A more sophisticated and structured method could also be implemented to guide askers in formulating their questions, particularly regarding paper recommendations. This would encourage them to conduct preliminary research before seeking help. For instance, askers could be prompted to provide detailed descriptions of their specific problem, summarize their findings to date, and list relevant papers they have already encountered. This process not only ensures that askers are actively engaged in their research but also helps experts understand the context and depth of the inquiry, allowing for more targeted and effective assistance. These enhancements aim to create a more balanced and efficient exchange of knowledge within the academic community.

*7.2.2 The Risk of De-anonymization.* Meronymity in academic settings poses a risk of unintended identity disclosure, especially in smaller fields or when specific details such as institution and field, and/or advisor's name are mentioned. Participants in the study expressed concerns about breaching their own or collaborators' privacy, leading to a cautious approach in their public disclosures of their identity signals. Yet, most of the time they were more open to sharing details with experts in private, leveraging the meronymity model's flexibility to control information revealed based on the level of their de-anonymization fears. Some were more comfortable sharing publicly, especially when inquiring about unfamiliar topics, believing their lack of knowledge was justified. The level of anonymity was generally acceptable, with the understanding that effort could lead to identity discovery, which some were okay with. Over time, some participants reported increased comfort with sharing identity information within the system. This change suggests initial reticence due to unfamiliarity, but growing ease over time.

The risk of de-anonymization is higher when combining various identity signals. To address this, a proactive approach could nudge users about the de-anonymization risks during the process of identity composition. This could be achieved by comparing the meronyms chosen by a user against a comprehensive database that identifies unique or potentially revealing meronyms. Additionally, the threat of de-anonymization increases with the use of algorithms capable of correlating an individual's activities across different social media platforms [37, 74]. Future work could examine ways to evaluate and mitigate these risks to protect consensual identity disclosure.

*7.2.3 The Rich-Get-Richer Effect.* The current meronymity model, which emphasizes connections with renowned experts and affiliations with prestigious institutions, risks creating a "rich get richer" effect. This model potentially privileges well-connected or institutionally advantaged students, allowing them to garner more attention and support, while junior students, isolated individuals, introverts, or those from marginalized groups may be overlooked. This disparity can widen the knowledge and career development gap between these groups.

To develop a more inclusive model, one idea we implemented, inspired by conversations with senior academic figures who expressed a strong desire to support junior students, is to highlight signals that identify askers as part of this less-privileged group, thereby motivating senior academics to assist. Additional affordances could be integrated into the system to showcase seniors' willingness to help, such as a "best helper award" or a karma-like score reflecting their engagement in answering questions. A related feature we implemented in LiTweeture highlighted professors who were willing to answer questions, an affordance that was well-received. These indicators could assist students in identifying responsive experts, enabling them to receive help without needing direct connections.

*7.2.4 Threats from Bad Actors.* It's important, when designing a social platform, to prevent its misuse for harassment or other anti-social behavior. Although LiTweeture introduced Endorsers and Helpers in part to *prevent* such behavior, the introduction also creates new channels to *support it*. For example, an adversarial endorser could endorse harassing posts from any number of adversarial posters. Since we permit endorsers to endorse meronymously, it would not be possible to identify the particular adversarial endorser *or* the posters they endorse, and all could escape consequences. We could fix this problem by requiring full identities for endorsers.

But as we saw in our interviews, some endorsers wished to be meronymous, so this would decrease the pool of willing endorsers.

## 7.3 Broader Applications of Meronymity

*7.3.1 Other Use Cases of Meronymity in Academia.* Although the current version of LiTweeture was designed for asking about paper recommendations, participants also found other uses for it. For example, one junior participant used LiTweeture to ask the CHI community about their opinions on the new Revise and Resubmit review process while only revealing that they had published at the CHI conference in the past. Another example was from a more senior participant who wanted to use LiTweeture to start discussions around specific papers. Another feature proposal was to use LiTweeture to organize *"ask-me-anything"* styled discussion with revealed partial identity to show expertise in a research area. Finally, while in our user study we focused on supporting junior members of the research community, our interview also pointed to marginalized groups as an important future work. For example, one faculty member said that *"As a woman in computer science I realized that many times my superior, more senior researchers, from my personal experience, can have a hard time believing what I'm doing and see me as an expert... [Sometimes] what I do is communicate in private channels where I'm not so worried about, I'm presenting myself in a certain way."*

*7.3.2 Deploying Meronymity for General-Purpose Discussion.* We believe our meronymity model can be generalized with further design refinements. In the academic context, as demonstrated in LiTweeture, identity verification relied on automatically generated signals from a trusted source, Semantic Scholar, focusing on publication history for paper recommendation inquiries. This approach can be applied to any other existing social media platform with basic functionality of posting content publicly and direct messaging. However, this definition of identity is less relevant outside academic contexts.

For broader applications, we propose a free-form self-description method for identity expression. This presents the challenge of verifying these self-declared identities. To address potential misrepresentations, such as someone falsely claiming to be a descendant of Einstein, we suggest a peer-verification model. Here, identities are confirmed by known associates on the platform, with each verified description marked accordingly and the verifier's identity disclosed for accountability.

Despite this, risks of endorsing false identities or misuse for harassment, spam, or misinformation remain. To mitigate these concerns, we advocate integrating this identity system with a content moderation model based on user endorsements, akin to the "squadbox" moderation model [58]. After a user posts using a meronymous identity, they reveal their identity to selected endorsers. If an endorser endorses the post, it becomes visible to their followers, thereby placing trust and accountability on the endorsers. Such a model prioritizes trustworthiness and accountability in the verification and dissemination of content on meronymous platforms.

## 8 LIMITATIONS

Our study faced several limitations. Firstly, the limited participant count of 13 led to an imbalance in participant distribution among conditions (due to drop outs) and a low volume of questions. This allowed participants to easily review all questions when selecting a question on LiTweeture to answer as part of their weekly task, focusing more on expertise than the poster's identity. While identity signals did motivate responses and help tailor answers, the small question pool may have diminished the influence of these signals in question selection, as noted by some participants. Future research should examine the impact of identity signals on a larger scale to gain more insights.

Additionally, the acquisition of Twitter by Elon Musk coincided with our system deployment, potentially affecting response rates and lowering participant reported satisfaction due to fewer received recommendations through LiTweeture.

Many participants found it challenging to identify three relevant experts for their questions. Implementing an expert matching model could significantly aid in connecting users with appropriate assistance.

While our system was effective for broader queries, the asynchronous nature of social media hindered deeper follow-up discussions about suggested work. To overcome this, we propose introducing anonymized synchronous chat features, enabling more in-depth conversations without identity disclosure. Additionally, allowing users to selectively reveal their identities to specific parties could facilitate collaborations, addressing a desire expressed by several participants.

## 9 CONCLUSION

In societies with pronounced social hierarchies, individuals of lower status often feel inhibited to voice their opinions, especially in front of their higher-status counterparts. This is evident in the academic community, a traditionally hierarchical setting, where the high stakes associated with one's reputation and social standing intensify the pressure of public discourse. This work introduces *meronymity* as a paradigm allowing individuals to selectively reveal facets of their identity in public conversations, aiming to strike a balance between anonymity and full disclosure. Using the academic community as a test-bed, we conducted a formative study to deepen our understanding of the dynamics of seeking and sharing research paper recommendations on social platforms like Twitter. Our findings revealed a notable hesitation to participate in public conversations, driven by concerns about personal reputation, fear of criticism, and potential judgments, especially from senior figures in the field. Interestingly, there was a distinct inclination towards partially anonymous interactions, which offer a balance between privacy and authenticity, enriching knowledge exchange. To operationalize these findings, we presented design objectives for facilitating more engaging and partially-anonymous online public communication. We developed a system called LiTweeture to incorporate these design goals. We recruited 13 participants to use LiTweeture during a one month field study. Our results indicated that meronymity indeed elevated the comfort levels of junior researchers in publicly seeking advice, thus broadening their research horizons. Additionally, providing selective identity cues added context and credibility to discussions, enhancing the overall quality of interactions while concealing identity. Our research highlights the transformative potential of meronymous communication in online platforms, particularly in contexts with strong hierarchical dynamics. The challenge ahead lies in fine-tuning such systems to best serve individual and contextual needs.

## ACKNOWLEDGMENTS

We extend our gratitude to Farnaz Jahanbakhsh for her valuable input and the Semantic Scholar team for their insightful discussions and feedback. We also appreciate the constructive comments from our anonymous reviewers. Lastly, we acknowledge our user study participants for their essential role in this work.

# A  ENDORSEMENT INTERFACE

# B  ASKING INTERFACE

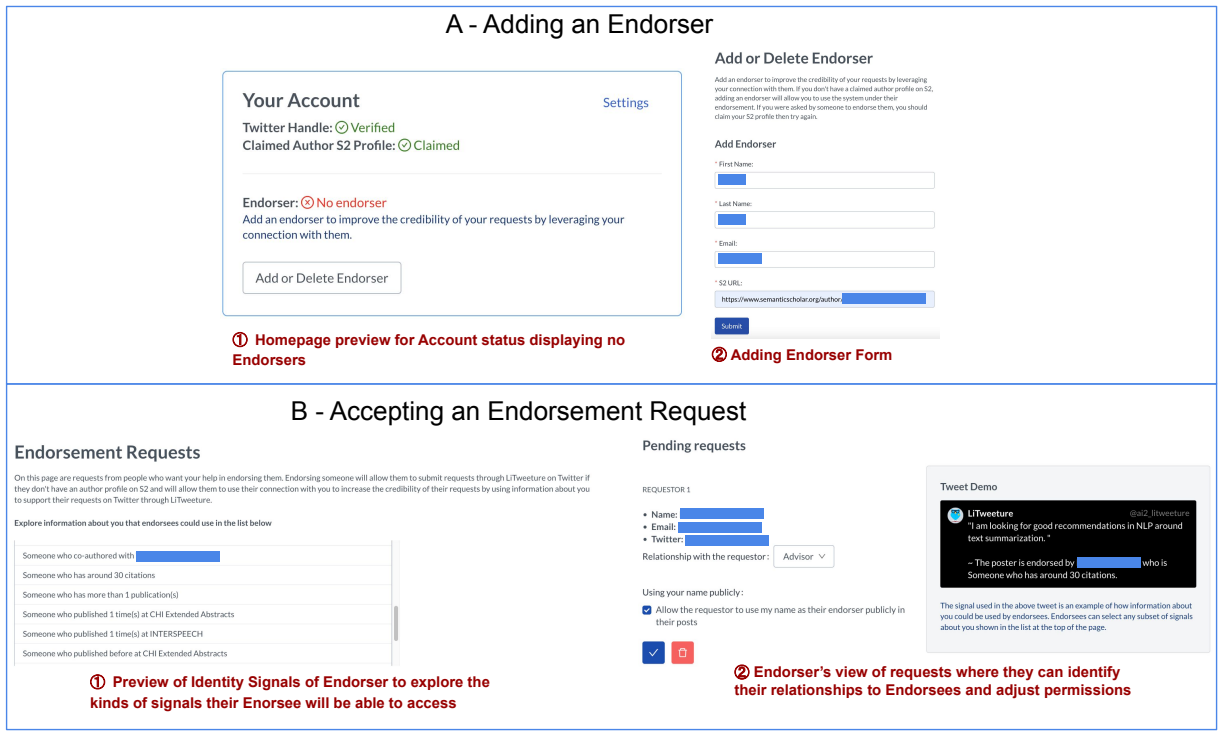

Figure 6: The Endorsement Interface in LiTweeture. Part A of this figure showcases the process and interface for adding an endorser. The interface allows for the provision of the endorser's Name, contact email and Semantic Scholar author page. Part B illustrates how endorsers can accept endorsement requests. Endorsers can specify their relationship to the endorsee and control the disclosure of their identity by endorsee. The system emphasizes informed consent with previews of the identity signals shared in an example tweet, ensuring both parties can manage the endorsement relationship transparently.

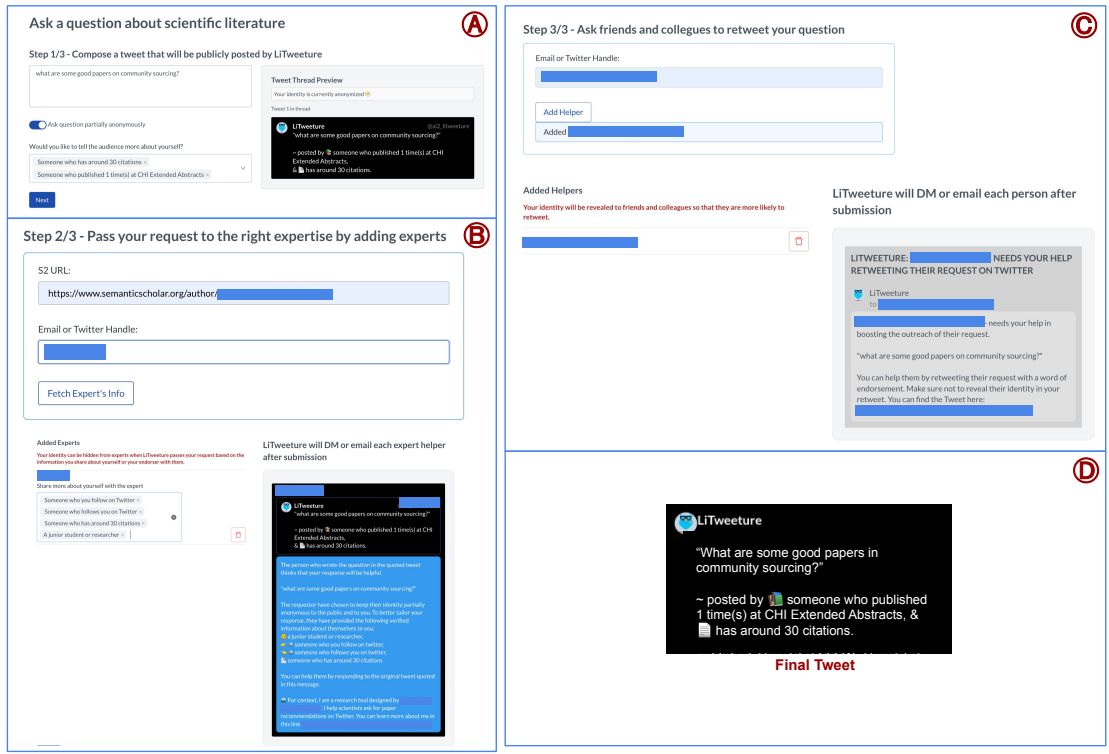

Figure 7: The Question-Asking Interface of LiTweeture. This figure displays the sequence of steps a user takes to ask a question through LiTweeture. Step 1 (A) shows how users can compose their query, compose their public meronym and preview the tweet as it will appear when posted. In Step 2 (B), users select experts for their inquiry by providing contact details and author profiles, compose their personalized meronyms to each expert, and preview the expert invitations as they will appear to experts. Step 3 (C) allows users to add helpers who can amplify their request, and preview the helper invitations as they will appear to helpers. The final tweet (D) depicts the anonymized public post as it will be displayed on Twitter.